\documentclass[manuscript]{aastex63}
\usepackage[]{graphicx}
\usepackage{float}
\usepackage[figuresleft]{rotating}
\usepackage{amsmath}
\usepackage{mathtools}
\usepackage{comment}
\usepackage{breakurl}
\usepackage{lineno}

\usepackage{multirow}
\usepackage{threeparttable}

\newcommand{\Msun}{$M_\odot$}

\newcommand{\Lsun}{$L_\odot$}

\newcommand{\co}{C$^{18}$O}
\newcommand{\CO}{\co}
\newcommand{\SOO}{SO$_2$}
\newcommand{\thso}{$^{34}$SO}
\newcommand{\SO}{\thso}
\newcommand{\HCO}{H$^{13}$CO$^+$}

\newcommand{\HCOmain}{HCO$^+$}
\newcommand{\DCO}{DCO$^+$}

\newcommand{\hydro}{H$_2$}

\newcommand{\kms}{km~s$^{-1}$}
\newcommand{\kmps}{\kms}
\newcommand{\mjybeam}{mJy~beam$^{-1}$}
\newcommand{\mjypb}{\mjybeam}
\newcommand{\ujybeam}{$\mu$Jy~beam$^{-1}$}
\newcommand{\ujypb}{\ujybeam}
\newcommand{\Trot}{$T_{\rm rot}$}
\newcommand{\Tkin}{$T_{\rm kin}$}
\newcommand{\cminvsq}{${\rm cm}^{-2}$} 
\newcommand{\cminvcb}{${\rm cm}^{-3}$}
\newcommand{\chisq}{$\chi^2$}

\newcommand{\soa}{$J_N=6_5-5_4$}
\newcommand{\thsoa}{$J_N=5_6-4_5$}
\newcommand{\sob}{$J_N=6_6-5_5$}

\newcommand{\coa}{$J=2-1$}
\newcommand{\csa}{$J=5-4$}
\newcommand{\sioa}{$J=5-4$}

\newcommand{\hcoa}{$J=3-2$}

\newcommand{\dcoa}{$J=3-2$}

\newcommand{\Bstar}{HD~147889}

\newcommand{\fieldcentposition}{($\alpha_{\rm ICRS}$, $\delta_{\rm ICRS}$) $=$ 
($16^{\rm h}27^{\rm m}09.\!\!^{\rm s}437$, $-24\degr37\arcmin19\farcs304$)}

\newcommand{\contpeakposition}{($\alpha_{\rm ICRS}$, $\delta_{\rm ICRS}$) $=$ 
($16^{\rm h}27^{\rm m}09.\!\!^{\rm s}41737\pm0.\!\!^{\rm s}00007$, $-24\degr37\arcmin19\farcs3104\pm0\farcs0007$)}

\newcommand{\posAD}{Positions~A$-$D}

\newcommand{\posAE}{Positions~A$-$E}

\newcommand{\posBD}{Positions~B$-$D}

\newcommand{\CPposition}{($\alpha_{\rm ICRS}$, $\delta_{\rm ICRS}$) $=$ 
($16^{\rm h}27^{\rm m}09.\!\!^{\rm s}417$, $-24\degr37\arcmin19\farcs310$)}

\newcommand{\noteCoordsAEwithoutFig}{
Coordinates of the \posAE\ are summarized in Table~\ref{tb:coords}}

\newcommand{\noteCoordsADwithoutFig}{
Coordinates of the \posAD\ are summarized in Table~\ref{tb:coords}}

\newcommand{\noteAverage}{Each value is obtained as the average over the circular region with a diameter of 1\arcsec\ centered at each of \posAD}

\newcommand{\noteRMS}[1]{{\bff $\sigma$ of {#1} is the rms noise level}} 

\newcommand{\iffigure}{\iftrue}
\newcommand{\dirfig}{} 
\newcommand{\ifshortauthorslist}{\iffalse}
\newcommand{\bff}{}

\newcommand{\yo}[1]{\textcolor{magenta}{\bff #1}}

\shorttitle{}
\shortauthors{}
\graphicspath{{./}{figures/}}

\NewPageAfterKeywords
\begin{document}

\title{Evidence for Jet/Outflow Shocks Heating the Environment Around the Class I Protostellar Source Elias 29: FAUST XXI} 

\author{Yoko Oya}
\affiliation{Center for Gravitational Physics, Yukawa Institute for Theoretical Physics, Kyoto University, Oiwake-cho, Kitashirakawa, Sakyo-ku, Kyoto-shi, Kyoto-fu 606-8502, Japan}

\author{Eri Saiga}
\affiliation{Department of Physics, The University of Tokyo, 7-3-1, Hongo, Bunkyo-ku, Tokyo 113-0033, Japan}

\author{Anna Miotello}
\affiliation{European Southern Observatory, Karl-Schwarzschild Str. 2, 85748 Garching bei M\"{u}nchen, Germany}

\author{Maria Koutoulaki}
\affiliation{School of Physics \& Astronomy, University of Leeds, Woodhouse Lane, LS2 9JT Leeds, UK}


\author{Doug Johnstone}
\affiliation{NRC Herzberg Astronomy and Astrophysics, 5071 West Saanich Road, Victoria, BC, V9E 2E7, Canada}
\affiliation{Department of Physics and Astronomy, University of Victoria, Victoria, BC, V8P 5C2, Canada}

\author{Cecilia Ceccarelli}
\affiliation{Universite Grenoble Alpes, CNRS, IPAG, 38000 Grenoble, France}

\author{Claire J. Chandler}
\affiliation{National Radio Astronomy Observatory, PO Box O, Socorro, NM 87801, USA}

\author{Claudio Codella}
\affiliation{INAF, Osservatorio Astrofisico di Arcetri, Largo E. Fermi 5, I-50125, Firenze, Italy}
\affiliation{Universite Grenoble Alpes, CNRS, IPAG, 38000 Grenoble, France}

\author{Nami Sakai}
\affiliation{RIKEN Cluster for Pioneering Research, 2-1, Hirosawa, Wako-shi, Saitama 351-0198, Japan}


\author{Eleonora Bianchi}
\affiliation{INAF, Osservatorio Astrofisico di Arcetri, Largo E. Fermi 5, I-50125, Firenze, Italy}

\author{Mathilde Bouvier}
\affiliation{Leiden Observatory, Leiden University, P.O. Box 9513, 2300 RA Leiden, The Netherlands}

\author{Steven Charnley}
\affiliation{Instituto de Radioastronom\'{i}a y Astrof\'{i}sica, Universidad Nacional Aut\'{o}noma de M\'{e}xico, 58090 Morelia, Mexico}

\author{Nicolas Cuello}
\affiliation{Universite Grenoble Alpes, CNRS, IPAG, 38000 Grenoble, France}

\author{Marta De Simone}
\affiliation{European Southern Observatory, Karl-Schwarzschild Str. 2, 85748 Garching bei M\"{u}nchen, Germany}
\affiliation{INAF, Osservatorio Astrofisico di Arcetri, Largo E. Fermi 5, I-50125, Firenze, Italy}

\author{Logan Francis}
\affiliation{Leiden Observatory, Leiden University, PO Box 9513, 2300 RA Leiden, The Netherlands}

\author{Tomoyuki Hanawa}
\affiliation{Center for Frontier Science, Chiba University, 1-33 Yayoi-cho, Inage-ku, Chiba 263-8522, Japan}

\author{Izaskun Jim\'enez-Serra}
\affiliation{Centro de Astrobiolog\'{\i}a (CAB), CSIC-INTA, Ctra. de Ajalvir, km 4, 28850, Torrej\'on de Ardoz, Spain}

\author{Laurent Loinard}
\affiliation{Instituto de Radioastronom\'{i}a y Astrof\'{i}sica, Universidad Nacional Auton\'{o}ma de M\'{e}xico, Apartado Postal 3-72, Morelia 58090, Michoac\'{a}n, Mexico}
\affiliation{Black Hole Initiative at Harvard University, 20 Garden Street, Cambridge, MA 02138, USA}
\affiliation{David Rockefeller Center for Latin American Studies, Harvard University, 1730 Cambridge Street, Cambridge, MA 02138, USA}

\author{Francois Menard}
\affiliation{Universite Grenoble Alpes, CNRS, IPAG, 38000 Grenoble, France}

\author{Giovanni Sabatini} 
\affiliation{INAF, Osservatorio Astrofisico di Arcetri, Largo E. Fermi 5, I-50125, Firenze, Italy}

\author{Charlotte Vastel}
\affiliation{IRAP, Universit\'{e} de Toulouse, CNRS, CNES, UPS, Toulouse, France}

\author{Ziwei Zhang}
\affiliation{RIKEN Cluster for Pioneering Research, 2-1, Hirosawa, Wako-shi, Saitama 351-0198, Japan}


\author{Yuri Aikawa}
\affiliation{Department of Astronomy, The University of Tokyo, 7-3-1 Hongo, Bunkyo-ku, Tokyo 113-0033, Japan}

\author{Felipe O. Alves}
\affiliation{Center for Astrochemical Studies, Max-Planck-Institut f\"{u}r extraterrestrische Physik (MPE), Gie$\beta$enbachstr. 1, D-85741 Garching, Germany}

\author{Nadia Balucani}
\affiliation{Department of Chemistry, Biology, and Biotechnology, The University of Perugia, Via Elce di Sotto 8, 06123 Perugia, Italy}

\author{Gemma Busquet}
\affiliation{Universite Grenoble Alpes, CNRS, IPAG, 38000 Grenoble, France}
\affiliation{Departament de F\'{i}sica Qu\'{a}ntica i Astrof\'{i}sica, Institut de Ci\'{e}ncies del Cosmos, Universitat de Barcelona (IEEC-UB), Mart\'{i} i Franqu\'{e}s, 1, 08028, Barcelona, Catalunya, Spain}

\author{Paola Caselli}
\affiliation{Center for Astrochemical Studies, Max-Planck-Institut f\"{u}r extraterrestrische Physik (MPE), Gie$\beta$enbachstr. 1, D-85741 Garching, Germany}

\author{Emmanuel Caux}
\affiliation{IRAP, Universit\'{e} de Toulouse, CNRS, CNES, UPS, Toulouse, France}

\author{Spandan Choudhury}
\affiliation{Center for Astrochemical Studies, Max-Planck-Institut f\"{u}r extraterrestrische Physik (MPE), Gie$\beta$enbachstr. 1, D-85741 Garching, Germany}

\author{Francois Dulieu}
\affiliation{CY Cergy Paris Universit\'{e}, Sorbonne Universit\'{e}, Observatoire de Paris, PSL University, CNRS, LERMA, F-95000, Cergy, France}

\author{Aurora Dur\'{a}n}
\affiliation{Instituto de Radioastronom\'{i}a y Astrof\'{i}sica , Universidad Nacional Aut\'{o}noma de M\'{e}xico, A.P. 3-72 (Xangari), 8701, Morelia, Mexico}

\author{Lucy Evans}
\affiliation{IRAP, Universit\'{e} de Toulouse, CNRS, CNES, UPS, Toulouse, France}
\affiliation{INAF, Osservatorio Astrofisico di Arcetri, Largo E. Fermi 5, I-50125, Firenze, Italy}


\author{Davide Fedele}
\affiliation{INAF, Osservatorio Astrofisico di Arcetri, Largo E. Fermi 5, I-50125, Firenze, Italy}
\affiliation{INAF, Osservatorio Astrofisico di Torino, Via Osservatorio 20, 10025, Pino Torinese, Italy}

\author{Siyi Feng}
\affiliation{Department of Astronomy, Xiamen University, Xiamen, Fujian 361005, P. R. China}

\author{Francesco Fontani}
\affiliation{INAF, Osservatorio Astrofisico di Arcetri, Largo E. Fermi 5, I-50125, Firenze, Italy}
\affiliation{Center for Astrochemical Studies, Max-Planck-Institut f\"{u}r extraterrestrische Physik (MPE), Gie$\beta$enbachstr. 1, D-85741 Garching, Germany}

\author{Tetsuya Hama}
\affiliation{Komaba Institute for Science, The University of Tokyo, 3-8-1 Komaba, Meguro, Tokyo 153-8902, Japan}
\affiliation{Department of Basic Science, The University of Tokyo, 3-8-1 Komaba, Meguro, Tokyo 153-8902, Japan}

\author{Eric Herbst}
\affiliation{Department of Chemistry, University of Virginia, McCormick Road, PO Box 400319, Charlottesville, VA 22904, USA}

\author{Shingo Hirano}
\affiliation{Department of Astronomy, The University of Tokyo, 7-3-1 Hongo, Bunkyo-ku, Tokyo 113-0033, Japan}

\author{Tomoya Hirota}
\affiliation{National Astronomical Observatory of Japan, Osawa 2-21-1, Mitaka-shi, Tokyo 181-8588, Japan}
\affiliation{Department of Astronomical Sciences, SOKENDAI (The Graduate University for Advanced Studies), 2-21-1 Osawa, Mitaka-shi, Tokyo 181-8588, Japan} 


\author{Andrea Isella}
\affiliation{Department of Physics and Astronomy, Rice University, 6100 Main Street, MS-108, Houston, TX 77005, USA}

\author{Claudine Kahane}
\affiliation{Universite Grenoble Alpes, CNRS, IPAG, 38000 Grenoble, France}

\author{Bertrand Lefloch}
\affiliation{Universite Grenoble Alpes, CNRS, IPAG, 38000 Grenoble, France}

\author{Romane Le Gal}
\affiliation{Universite Grenoble Alpes, CNRS, IPAG, 38000 Grenoble, France}
\affiliation{Institut de Radioastronomie Millim\'{e}trique, 38406 Saint-Martin d'H$\grave{e}$res, France}

\author{Hauyu Baobab Liu}
\affiliation{Institute of Astronomy and Astrophysics, Academia Sinica, 11F of Astronomy-Mathematics Building, AS/NTU No.1, Sec. 4, Roosevelt Rd., Taipei 10617, Taiwan, R.O.C.}

\author{Ana L\'{o}pez-Sepulcre}
\affiliation{Universite Grenoble Alpes, CNRS, IPAG, 38000 Grenoble, France}
\affiliation{Institut de Radioastronomie Millim\'{e}trique, 38406 Saint-Martin d'H$\grave{e}$res, France}

\author{Luke T. Maud}
\affiliation{European Southern Observatory, Karl-Schwarzschild Str. 2, 85748 Garching bei M\"{u}nchen, Germany}

\author{Mar\'{i}a Jos\'{e} Maureira}
\affiliation{Center for Astrochemical Studies, Max-Planck-Institut f\"{u}r extraterrestrische Physik (MPE), Gie$\beta$enbachstr. 1, D-85741 Garching, Germany}

\author{Seyma Mercimek}
\affiliation{INAF, Osservatorio Astrofisico di Arcetri, Largo E. Fermi 5, I-50125, Firenze, Italy}
\affiliation{Universit$\grave{a}$ degli Studi di Firenze, Dipartimento di Fisica e Astronomia, via G. Sansone 1, 50019 Sesto Fiorentino, Italy}


\author{George Moellenbrock}
\affiliation{National Radio Astronomy Observatory, PO Box O, Socorro, NM 87801, USA}

\author{Shoji Mori}
\affiliation{Department of Astronomy, The University of Tokyo, 7-3-1, Hongo, Bunkyo-ku, Tokyo 113-0033, Japan}



\author{Hideko Nomura}
\affiliation{Division of Science, National Astronomical Observatory of Japan, 2-21-1 Osawa, Mitaka, Tokyo 181-8588, Japan}

\author{Yasuhiro Oba}
\affiliation{Institute of Low Temperature Science, Hokkaido University, N19W8, Kita-ku, Sapporo, Hokkaido 060-0819, Japan}

\author{Ross O'Donoghue}
\affiliation{Department of Physics and Astronomy, University College London, Gower Street, London, WC1E 6BT, UK}

\author{Satoshi Ohashi}
\affiliation{National Astronomical Observatory of Japan, Osawa 2-21-1, Mitaka-shi, Tokyo 181-8588, Japan}

\author{Yuki Okoda}
\affiliation{RIKEN Cluster for Pioneering Research, 2-1, Hirosawa, Wako-shi, Saitama 351-0198, Japan}

\author{Juan Ospina-Zamudio}
\affiliation{Universite Grenoble Alpes, CNRS, IPAG, 38000 Grenoble, France}


\author{Jaime Pineda}
\affiliation{Center for Astrochemical Studies, Max-Planck-Institut f\"{u}r extraterrestrische Physik (MPE), Gie$\beta$enbachstr. 1, D-85741 Garching, Germany}

\author{Linda Podio}
\affiliation{INAF, Osservatorio Astrofisico di Arcetri, Largo E. Fermi 5, I-50125, Firenze, Italy}

\author{Albert Rimola}
\affiliation{Departament de Qu\'{i}mica, Universitat Aut$\grave{o}$noma de Barcelona, 08193 Bellaterra, Spain}

\author{Takeshi Sakai}
\affiliation{Graduate School of Informatics and Engineering, The University of Electro-Communications, Chofu, Tokyo 182-8585, Japan}

\author{Dominique Segura-Cox}
\affiliation{Center for Astrochemical Studies, Max-Planck-Institut f\"{u}r extraterrestrische Physik (MPE), Gie$\beta$enbachstr. 1, D-85741 Garching, Germany}

\author{Yancy Shirley}
\affiliation{Steward Observatory, 933 N Cherry Ave., Tucson, AZ 85721 USA}

\author{Brian Svoboda}
\affiliation{National Radio Astronomy Observatory, PO Box O, Socorro, NM 87801, USA}
\altaffiliation{Jansky Fellow of the National Radio Astronomy Observatory.}


\author{Leonardo Testi}
\affiliation{European Southern Observatory, Karl-Schwarzschild Str. 2, 85748 Garching bei M\"{u}nchen, Germany}
\affiliation{INAF, Osservatorio Astrofisico di Arcetri, Largo E. Fermi 5, I-50125, Firenze, Italy}

\author{Serena Viti}
\affiliation{Leiden Observatory, Leiden University, PO Box 9513, NL02300 RA,  Leiden, The Netherlands}

\author{Naoki Watanabe}
\affiliation{Institute of Low Temperature Science, Hokkaido University, N19W8, Kita-ku, Sapporo, Hokkaido 060-0819, Japan}

\author{Yoshimasa Watanabe}
\affiliation{Materials Science and Engineering, College of Engineering, Shibaura Institute of Technology, 3-7-5 Toyosu, Koto-ku, Tokyo 135-8548, Japan}



\author{Yichen Zhang}
\affiliation{Department of Astronomy, Shanghai Jiao Tong University, 800 Dongchuan Rd., Minhang, Shanghai 200240, China} 


\author{Satoshi Yamamoto}
\affiliation{SOKENDAI (The Graduate University for Advanced Studies), Hayama-cho, Miura-gun, Kanagawa 240-0193, Japan}
\affiliation{Research Center for the Early Universe, The University of Tokyo, 7-3-1, Hongo, Bunkyo-ku, Tokyo 113-0033, Japan}

\ifshortauthorslist
\collaboration{4}{}
\fi
\correspondingauthor{Yoko Oya}
\email{yoko.oya@yukawa.kyoto-u.ac.jp}

\begin{abstract}
We have observed the late Class I protostellar source Elias~29 at a spatial-resolution of 70~au with the Atacama~Large~Millimeter/submillimeter~Array (ALMA) as part of the FAUST Large Program. We focus on the line emission of SO, while that of $^{34}$SO, C$^{18}$O, CS, SiO, H$^{13}$CO$^{+}$, and DCO$^{+}$ are used supplementally. The spatial distribution of the SO rotational temperature ($T_{\rm rot}$(SO)) is evaluated by using the intensity ratio of its two rotational excitation lines. Besides in the vicinity of the protostar, two hot spots are found at a distance of 500 au from the protostar; $T_{\rm rot}$(SO) locally rises to 53$^{+25}_{-15}$~K at the interaction point of the outflow and the southern ridge, and 72$^{+66}_{-29}$~K within the southeastern outflow probably due to a jet-driven bow shock. However, the SiO emission is not detected at these hot spots. It is likely that active gas accretion through the disk-like structure and onto the protostar still continues even at this evolved protostellar stage, at least sporadically, considering the outflow/jet activities and the possible infall motion previously reported. Interestingly, $T_{\rm rot}$(SO) is as high as 20$-$30~K even within the quiescent part of the southern ridge apart from the protostar by 500$-$1000~au without clear kinematic indication of current outflow/jet interactions. Such a warm condition is also supported by the low deuterium fractionation ratio of HCO$^+$ estimated by using the H$^{13}$CO$^{+}$ and DCO$^{+}$ lines. The B-type star HD147889 $\sim$0.5 pc away from Elias~29, previously suggested as a heating source for this region, is likely responsible for the warm condition of Elias~29. 
\end{abstract}

\section{Introduction} \label{sec:intro}

\subsection{Background} \label{sec:intro_bg}

A wide variety of planetary systems have been discovered in recent decades 
\citep[e.g.][]{Andrews et al.(2018), Oberg et al.(2021)}. 
The origin of their diversity, both in the physics and chemistry,
probably resides in the earliest history of the planetary system formation; 
namely, what happened during the protostellar phases. 
Thus, it is important to investigate this phase for thorough understanding of star and planet formation and its diversity. 
In this phase, gas accretion to the protostar and dissipation of surrounding gas are simultaneously ongoing, 
so that a parent core generally reveals a very complex structure. 
Recent ALMA observations have indeed delineated 
a detailed view of the complex physical and chemical nature of protostellar sources at a high angular resolution 
\citep[e.g.][]{Tokuda et al.(2014), van der Wiel et al.(2019), Okoda et al.(2021), Ohashi et al.(2022), Codella et al.(2024)}. 
In both gas accretion and dissipation, the outflow/jet plays a central role 
\citep[e.g.][]{Bachiller(1996), Bally(2016)}. 

First, outflows are known to be important mechanisms 
for removing angular momentum from the gas accreting onto the protostar, 
and therefore closely connected to the formation of the disk/envelope system and the growth of protostars 
\citep[e.g.][]{Blandford & Payne(1982), Tomisaka(2000), Machida et al.(2008), Hirota et al.(2017), Machida & Basu(2019)}. 
Second, 
the outflow plays a crucial role in the dissipation of the parent core, 
the details of which are still an open issue of theoretical study 
\citep[e.g.][]{Offner & Chaban(2017), Nakamura & Li(2014)}. 
The outflow evolution 
is thus important 
for both regulation and feedback processes in the formation of a star and the circumstellar environment. 
Hence, 
outflows have {\bff extensively been} studied by radio observations, particularly for early-stage protostars 
\citep[e.g.][]{Lee et al.(2000), Arce & Sargent(2006), Hatchell et al.(2007), Curtis et al.(2010), Matsushita et al.(2019), Nakamura et al.(2011) Serpens, Oya et al.(2014), Oya et al.(2015), Oya et al.(2018), Oya et al.(2018) 16293B, Oya et al.(2021), Omura et al.(2024)}. 
Furthermore, 
it is known that 
the chemical composition of a parent core is greatly affected by the outflow activity 
\citep[e.g.][]{Mikami et al.(1992), Bachiller & Perez Gutierrez(1997), Bachiller et al.(2001), Arce et al.(2008), Codella et al.(2010), Ceccarelli et al.(2010), Ceccarelli et al.(2017), Sugimura et al.(2011), Podio et al.(2017)}. 
Hence, the outflow stands for a key phenomenon 
to disentangle complex physical and chemical structure of protostellar cores. 

On the other hand, 
observational studies of outflows and {\bff their feedback} 
on the chemical evolution of evolved protostars at the late Class I stage 
are relatively sparse, especially at high angular resolution 
\citep[e.g.][]{Le Gal et al.(2020), Bianchi et al.(2022), Tanious et al.(2024)}. 
This is in part due to the lower mass accretion rates associated with evolved protostars, 
which generally drive less powerful and fainter outflows 
\citep[e.g.][]{Machida & Hosokawa(2013)}. 
Therefore, it is 
important to investigate additional sources at the late evolutionary stage, to attain a more thorough understanding of both the outflow evolution and its feedback to the parent core. 
Even if the outflow feature is faint, 
temperature structure as well as chemical structure around the protostar 
would tell us important information on the outflow activity. 
{\bff Moreover, 
accreted material at the late evolutionary stage remains in the disk 
and could have an outsize contribution to planets and comets. 
Thus, the nature of the late-accreting material would affect the chemical composition of comets; 
for instance, a supply of warm material could result in lower deuterium fractionation in comets than in the dense interstellar medium. 
Given these needs,}  
we included the relatively evolved protostellar source Elias~29
in our ALMA large program FAUST 
(Fifty AU Study of the chemistry in the disk/envelope system of solar-like protostars; 2018.1.01205.L; P.I.: S.~Yamamoto). 
FAUST aims to delineate physical and chemical structures of 13 protostellar sources 
at a spatial resolution of $\sim$50 au 
{\bff \citep[][]{Codella et al.(2021), Oya et al.(in prep.)}.} 

\subsection{Target Source: Elias~29} \label{sec:intro_target}
Elias~29, also known as WL 15 \citep{Wilking & Lada(1983)}, 
is a Class I protostar in the L1688 dark cloud within the Ophiuchus star-forming region 
\citep[$d=137$~pc;][]{Ortiz-Leon et al.(2017)}. 
\citet{Lommen et al.(2008)} reported a bolometric temperature and bolometric luminosity of 391 K and 13.6~\Lsun, respectively. 
The systemic velocity of the protostellar system is about 4~\kms\ \citep{Oya et al.(2019)}. 
Elias~29 is surrounded by many YSOs, 
making the environment of this source very complex \citep{Rocha & Pilling(2018)}. 

Previous observational studies of Elias~29 at $>$$10^4$~au scale 
have focused on its outflow structure 
\citep{Sekimoto et al.(1997), Ceccarelli et al.(2002), Ybarra et al.(2006), Bussmann et al.(2007), Nakamura et al.(2011) rhoOph, van der Marel et al.(2013)}. 
According to CO ($J=3-2$) observations with Heinrich Hertz Telescope by \citet{Bussmann et al.(2007)}, 
the outflow has an inverse S-like shape at a 4000~au scale. 
{\bff The S-like structure is presumably due to 
{\bff a temporal change in the outflow direction.} 
On a smaller scale around the protostar, 
\citet{Ceccarelli et al.(2002)} detected the outflow in CO ($J=6-5$) emission with the James Clerk Maxwell Telescope, 
noting that the blue-shifted and red-shifted lobes are on the west and east sides of the protostar, respectively.  
In addition, the protostellar jet was observed in the near-infrared by \citet{Ybarra et al.(2006)}, revealing H$_2$ knots produced by the precessing east-west jet,
consistent with the outflow morphology found by \citet{Ceccarelli et al.(2002)} and \citet{Bussmann et al.(2007)}.

The gas structures observed in Elias~29 at $<$$10^3$~au scale consist of two main components, 
according to the observations of various molecular lines by \citet{Lommen et al.(2008)} and \citet{Oya et al.(2019)}.
These are a compact 
central 
component associated to the protostar and an 
off-center 
dense gas clump 
(called the ``southern ridge'' hereafter). 
The southern ridge is $\sim$500~au ($\sim$4\arcsec) south of the protostar 
and extends along the northeast-southwest direction, 
{\bff and its origin is unclear at the moment \citep{Lommen et al.(2008), Oya et al.(2019)}.} 
\citet{Oya et al.(2019)} observed Elias~29 in SO and \SOO\ emission with ALMA, 
reporting rotational motion for the compact component associated to the protostar. 
The authors estimated the protostellar mass to be from 0.8 to 1.0 \Msun, assuming Keplerian motion, 
and a disk 
inclination angle from 65\degr\ to 90\degr\ (0\degr\ for face-on). 

The molecular gas within Elias~29 is reported to show peculiar chemical characteristics  \citep{Oya et al.(2019)}. 
According to their results, the SO and \SOO\ emission is very bright in the compact component around the protostar. 
The compact distribution of \SOO\ in this source has also been reported by \citet{Artur de la Villarmois et al.(2019)}. 
Such compact distributions of SO and \SOO\ have been reported for a few other sources 
(e.g. SVS 13 by \citealt{Codella et al.(2021) SVS} and \citealt{Bianchi et al.(2023)}; Oph IRS 44 by \citealt{Artur de la Villarmois et al.(2022)}). 
On the other hand, CS emission, 
which for systems with typical chemical characteristics is usually bright within the disk/envelope system
\citep[e.g.][]{Oya et al.(2015), Oya et al.(2017), Imai et al.(2016)}, 
is faint toward the continuum peak and is preferentially seen within the southern ridge. 
Both of the components are deficient in 
interstellar complex organic molecules (iCOMs), such as HCOOCH$_3$ and (CH$_3$)$_2$O, 
and unsaturated hydrocarbon molecules, such as CCH and c-C$_3$H$_2$. 
%
%
As a possible explanation for these chemical characteristics, 
\citet{Oya et al.(2019)} proposed a relatively high dust temperature ($\geq$20 K)
in the parent cloud; 
{\bff dust surface reactions to form the above molecular species would be insufficient 
because such a high dust temperature prevents 
their mother species (e.g., C and CO) from depletion onto dust grains 
\citep[see Section \ref{sec:warm_dco};][]{Oya(2022)}.} 
Indeed, \citet{Rocha & Pilling(2018)} reported that 
the parent cloud of Elias~29 is irradiated by two bright BV stars, S1 and \Bstar; 
their model calculation suggested that the gas temperature of the cloud should not have been below 20 K. 
The actual temperature in the environment around the protostar, however, has not been delineated observationally so far; 
{\bff therefore, 
we firstly delineate the temperature distribution in Elias~29 at a $10^{2-3}$~au scale in this project.}

{\bff 
In this paper, 
we focus on the distribution of the rotational temperature of SO by using two SO lines in ALMA Band 6. 
The details of the observations are given in Section~\ref{sec:obs}. 
Section~\ref{sec:res} describes the overall view of the observational results, 
including the findings of hot regions. 
The hot regions are also found to be characteristic in gas dynamics (Section~\ref{sec:feedback}); 
we discuss the feedback induced by the outflow (Section~\ref{sec:outflow}) and jet (Section~\ref{sec:jet}) activities in this source 
as the candidate cause for the local heating of the gas apart from the protostar. 
The mass accretion is discussed in Section~\ref{sec:accretion}. 
The chemical composition of this source is explored in Section~\ref{sec:chem}. 
The SO {\bff column density} 
is evaluated in Section~\ref{sec:abundance} 
and found not to be significantly enhanced by the shock chemistry. 
As a support for the warm environment of this source, 
the deuterium fractionation ratio of HCO$^+$ is discussed 
by using the \HCO\ and \DCO\ emissions in Section~\ref{sec:warm_dco}. 
Section~\ref{sec:summary} gives the summary of this paper. 
}

\section{Observation} \label{sec:obs}
With ALMA, we observed the field around Elias~29 
between 2018 October and 2020 March during Cycle 6 operation, 
as part of the ALMA large program FAUST (2018.1.01205.L). 
Spectral lines listed in Table~\ref{tb:molecule} were observed in Band 6 using two frequency ranges, 
Setup~1 (214.0$-$219.0~GHz, 229.0$-$234.05~GHz) and Setup~2 (242.5$-$247.5~GHz, 257.5$-$262.5~GHz). 
In each spectral setup, we used the 12-m array data with two different antenna configurations, 
C43-4 and C43-1, 
as well as the 7-m array data of the Atacama Compact Array (ACA/Morita Array). 
{\bff Both Setups~1 and 2 were observed by 12~spectral windows 
with the bandwidth and the frequency resolution of 59~MHz and 122~kHz 
(0.15~\kmps\ at 250~GHz), respectively, 
and one with the bandwidth and the frequency resolution of 1875~MHz and 1.129~MHz 
(1.4~\kmps\ at 250~GHz), respectively.} 
The basic parameters of the observations including calibrator sources are summarized in Table~\ref{tb:ObP}. 
The field center of the observations was \fieldcentposition. 
The {\bff system} temperature ($T_{\rm sys}$) 
was typically 
70$-$130 K during the observations. 
The data were reduced using the Common Astronomy Software Applications (CASA) package 
\citep{CASA Team et al.(2022)} 
utilizing a modified version of the ALMA calibration pipeline based on v.5.6.1-8.el7
and an additional in-house calibration routine to correct for the $T_{\rm sys}$ and spectral line data normalization\footnote{\url{https://faust-imaging.readthedocs.io/en/latest/}}. 
Self-calibration was carried out using line-free continuum emission for each configuration.
Details of the self-calibration process are described by \citet{Imai et al.(2022)}. 
The visibility data with the three different configurations were combined in the UV plane after the self-calibration. 
The absolute accuracy of the flux calibration was 10\% \citep{Warmels et al.(2018)}. 
Images of \CO\ (\coa), SO (\soa), SO (\sob), \SO\ (\thsoa), CS (\csa), SiO (\sioa), \HCO\ (\hcoa), and \DCO\ (\dcoa) were obtained 
with the CLEAN algorithm of CASA (v.5.7.2-4) by employing Briggs weighting with a robustness parameter of 0.5. 
The beam size of the line emission is about $0\farcs5 \times 0\farcs4$ 
by combining the data with all the antenna configurations (C43-4, C43-1, and ACA) 
as summarized in Table~\ref{tb:molecule}. 
{\bff The maximum recoverable scale\footnote{\bff See also Table 7.1 in ALMA Cycle 6 Technical Handbook \citep[][\url{https://almascience.nao.ac.jp/documents-and-tools/cycle6/alma-technical-handbook\#page=89}]{Warmels et al.(2018)}.} 
of the observation 
with the ACA is 30\farcs3 according to the QA2 report. 
The images are corrected for the primary beam attenuation.}

\section{Results} \label{sec:res}

\subsection{Overall Morphology} \label{sec:res_morphology}
Figure~\ref{fig:continuum} shows the 1.2 mm continuum map of Elias~29. 
The continuum emission has its brightest peak at the protostar position. 
The continuum peak position is derived to be \contpeakposition\ with 
a peak flux density of $14.62 \pm 0.06$ mJy beam$^{-1}$ 
based on 2-dimensional Gaussian fitting task {\tt imfit} with CASA for the continuum image. 
The convolved and deconvolved sizes of the continuum emission are 
$(0\farcs550 \pm 0\farcs003) \times (0\farcs404 \pm 0\farcs001)$ and $(0\farcs165 \pm 0\farcs009) \times (0\farcs152 \pm 0\farcs005)$, respectively, 
where the synthesized beam size is $0\farcs52\times0\farcs37$ (PA $=$ $-70.\!\!\degr0$). 
{\bff Besides this intensity peak, 
we found a weak north-south extension with a moderate emission peak 
{\bff of 0.63~\mjypb} 
at 600~au southeast of the protostar, 
{\bff which belongs to the southern ridge component.} 
Although this weak emission was not detected in \citet{Oya et al.(2019)}, 
it was detected in this project 
likely thanks to both the higher sensitivity and the wider UV coverage from the inclusion of an ACA configuration. 
Scientific examination based on the continuum emission,} 
including the 
{\bff 3 mm data taken as part of the FAUST program} 
as well as the 1.2 mm data used here, 
will be reported separately.

Figure~\ref{fig:mom0-8} shows the integrated and peak intensity maps of the SO (\soa), CS (\csa), and \CO\ (\coa) lines. 
{\bff The southern ridge component is bright in all the three species, 
while the continuum peak position is bright only in the SO emission. 
{\bff The} distribution of SO and CS are confirmed to be consistent with that previously reported by \citet{Oya et al.(2019)} 
with a higher spatial resolution and sensitivity of our data.}  
The distribution of \CO\ is 
found to be 
extended over the field of view and relatively weak at the continuum peak. 
{\bff {\bff The} 
\CO\ emission is generally more extended than the SO emission. 
{\bff Hence,} 
the former can be more filtered out by the interferometer, 
{\bff although the maximum recoverable scale of 30\farcs3 
will contribute to recover the extended emission.} 
{\bff Even if} the \CO\ emission 
{\bff were resolved out for the systemic velocity component
due to contamination of the extended ambient component, 
the} high velocity-shift {\bff components originating from}
the spin-up velocity structure {\bff near} the protostar 
\citep{Hartmann(2008), Ohashi et al.(2014), Oya et al.(2022)} 
{\bff should be detected as seen for the SO lines 
(Figure~\ref{fig:spectra-SO}; {\bff see Section~\ref{sec:res_temp}).}  
In addition, a compact distribution should also {\bff be} seen 
toward the protostar in the peak intensity map. 
A lack of these features (Figures~\ref{fig:mom0-8}f,~\ref{fig:spectra-SO}) indicates 
that the \CO\ emission is not resolved out 
significantly but} 
is really weak around the protostar. 
The peak intensity map of the \CO\ emission seems to have voids towards the northwest and southeast of the continuum peak position. 
This observed structure is likely related to the cavity produced by the outflow extending along the northwest-southeast direction 
\citep{Ceccarelli et al.(2002),Ybarra et al.(2006), Bussmann et al.(2007)}, 
as discussed further in Section \ref{sec:outflow}. 
{\bff We also note weak arc-like structures in the peak intensity map of the SO emission (Figure~\ref{fig:mom0-8}d) 
{\bff with the intensity up to 51~\mjypb\ detected by $\sim$5$\sigma$, 
where \noteRMS{10~\mjypb}}; 
the features on the eastern and western sides may trace the southeast and northwest outflow cavity walls, respectively (See also Section~\ref{sec:outflow}). 
Meanwhile, that on the southwestern side morphologically seems to be related 
to the southern ridge component.} 

\subsection{Temperature Distribution} \label{sec:res_temp}
In this section we investigate the distribution of the rotational temperature of SO by using its two observed transitions 
{\bff (\soa, $E_{\rm u1} = 35$ K; \sob, $E_{\rm u2} = 57$ K).} 
{\bff Assuming the LTE (local thermodynamic equilibrium) and optically thin conditions, 
the rotational temperature (\Trot) of SO is derived from the ratio of the integrated intensities of its two lines} 
according to the following equation: 
\begin{equation}
\frac{W_2}{W_1} = \frac{\nu_1 (S\mu^2)_2}{\nu_2 (S\mu^2)_1} \exp \left( - \frac{E_{\rm u2} - E_{\rm u1}}{k_{\rm B} T_{\rm rot}} \right), \label{eq:lte}
\end{equation}
where $W$ is the observed integrated intensity, 
$\nu$ the frequency of the transition, 
$S$ the line strength, 
$\mu$ the dipole moment responsible for the transition, 
and $E_{\rm u}$ the upper-state energy. 
{\bff The $\nu$, $S\mu^2$, and $E_{\rm u}$ values for each line 
are listed in Table~\ref{tb:molecule}, 
which are taken from CDMS 
\citep[The Cologne Database for Molecular Spectroscopy;][]{Muller et al.(2005), Endres et al.(2016)}.} 
Suffixes, 1 and 2, represent 
{\bff the \soa\ and \sob\ transitions,} 
respectively. 
{\bff Figure~\ref{fig:ratio-Trot_SO}(a) shows the map of the ratio of the integrated intensities of the two SO lines, 
and Figure~\ref{fig:ratio-Trot_SO}(b) shows the map of the evaluated \Trot\ of SO.}

The \Trot\ is mostly 20$-$30 K in the southern ridge component, 
{\bff except for a few small localized spots.} 
{\bff We} found three regions with a high rotational temperature in Figure~\ref{fig:ratio-Trot_SO}(b); 
the component around the continuum peak position labeled as `C.P.', 
the position at 500 au south of the protostar labeled as `Position~A' in the southern ridge component, 
and that at 500 au east of the protostar labeled as `Position~E'. 
{\bff Here, we specifically investigate 
the rotational temperature for these three positions 
and additional three positions in the southern ridge (Positions~B,~C,~and~D).} 
The coordinates of these positions are summarized in Table~\ref{tb:coords}. 
{\bff These positions are indicated on the maps in Figure~\ref{fig:ratio-Trot_SO}. 
\posAD\ are taken to be aligned on the white line along the extension of the southern ridge component in Figure~\ref{fig:ratio-Trot_SO}(a). 
Their central positions are separated by 2\arcsec\ from each other with an exception, 
the position between Position~A and Position~B, 
due to a complex velocity structure.} 
{\bff Figure~\ref{fig:spectra-SO} shows 
the spectra of the SO lines observed at \posAE\ and C.P..} 

The observed line parameters of SO for these positions are summarized in Table~\ref{tb:SOparams}. 
{\bff Here, the spectra at \posAD\ are 
averaged over a circular region with a diameter of 1\arcsec\ to increase the S/N ratio, 
while those at Position~E are taken at just 1~pixel to trace the compact structure of the bow-shocked region.} 
{\bff The line profiles are generally asymmetric 
and deviate from the Gaussian shape (Figure~\ref{fig:spectra-SO}). 
Thus, we do not employ the Gaussian fitting 
to derive the line parameters except for Position~E. 
For instance, 
the peak intensity of each line is evaluated from the peak value 
of the line profile (see the footnote of Table~\ref{tb:SOparams} for details).  
}

From the {\bff integrated} intensities of the two SO lines, the rotational temperature of SO 
{\bff at \posAE\ and the continuum peak} are derived by using Equation~(\ref{eq:lte}), 
as listed in Table~\ref{tb:temp-abundance_SO}. 
{\bff Since we use the integrated intensities in this analysis, 
the results do not suffer from the asymmetry of the line profile.} 
We discuss a possible cause for the high temperature 
at Positions~A (53$^{+25}_{-15}$~K) and E (72$^{+66}_{-29}$~K)  
in Sections \ref{sec:outflow} and \ref{sec:jet}, respectively. 
{\bff Although there is also another candidate spot with a high rotational temperature ($>60$~K) on the southern side of `Position~C', 
this result severely suffers from the error for the weak SO intensities; 
for instance, the rotational temperature is reduced to be 28~K 
by accounting for the 3$\sigma$ error (45~\mjypb~\kmps) of the SO (\sob) line {\bff integrated} intensity.} 

Since the evaluation of \Trot\ may suffer from the simple assumptions of the LTE and optically thin conditions, 
we confirmed its validity using a non-LTE analysis 
{\bff for \posAE. 
We used the two SO lines data and one \SO\ line data 
to constrain the three free parameters 
(the gas kinetic temperature, the SO column density, and the \hydro\ number density) 
by the \chisq\ method. 
We employed the photon-escaping probability for 
a static, spherically symmetric, and homogeneous medium 
\citep{Osterbrock & Ferland(2006), van der Tak et al.(2007)}. 
The spectra of the SO and \SO\ lines used for the non-LTE analysis 
are shown in Figure~\ref{fig:spectra-SO} for \posAE, 
whose parameters are summarized in Table~\ref{tb:SOparams}. 
Further details for the non-LTE analysis are} 
described in Appendix \ref{sec:appendix}. 

\section{Outflow/Jet Feedback as Candidate Causes for the Local Heating} \label{sec:feedback} 
\subsection{Temperature Enhancement due to Outflow Interactions} \label{sec:outflow}
In this section, we discuss the region's interaction with the outflow as the 
potential 
cause for the high temperature at Position~A. 
Figure~\ref{fig:mom0-outflow} shows {\bff the velocity channel maps} 
of the \CO\ and SO emissions. 
Although the overall structure is complex, 
we find a part of a parabolic feature extending over 20\arcsec\ ($\sim$2700 au) oriented toward southeast from the protostar 
{\bff in Figures~\ref{fig:mom0-outflow}(a, b, d).} 
{\bff We see a void of the \CO\ emission 
in its velocity channel map of $+5.4$~\kmps\ (Figure~\ref{fig:mom0-outflow}b) 
inside the parabolic shape on the northwestern side of the protostar, 
while such a feature is not evident in the SO emission (Figure~\ref{fig:mom0-outflow}e). 
This feature is seen probably because 
the \CO\ emission traces the cavity wall of the northwestern outflow lobe.} 

The outflow {\bff of} this source was previously reported based on the single-dish CO ($J=6-5$) observation by \citet{Ceccarelli et al.(2002)}, 
where the outflow axis is extending along the east-west direction on 30\arcsec\ scales. 
\citet{Bussmann et al.(2007)} reported the inverse S-shaped structure of the outflow; 
the outflow axis is along the southeast-northwest direction on 100\arcsec\ scales, 
while it is tilted to be along the south-north direction on larger scales. 
These outflow directions are consistent with the shocked knots seen in the \hydro\ infrared emission-line image reported by \citet{Ybarra et al.(2006)}. 
Considering these previous reports, 
the parabolic and void features found in Figure~\ref{fig:mom0-outflow} likely represent the outflow cavities. 

To explore the outflow-ridge interaction feature {\bff at Position A,}  
we produced a position-velocity (PV) diagram of the SO emission along the southern ridge (Figure~\ref{fig:SO-southernridge-PV}). 
The position axis of the PV diagram is taken to pass through the crossing point of the outflow cavity and the southern ridge, 
as shown in Figure~\ref{fig:mom0-outflow}(d). 
The spatially-extended component traces the southern ridge at the systemic velocity of $\sim$$+4.5$~\kmps. 
Moreover, 
a bright component is locally seen at Position~A 
with velocity ranging from $+4.5$~\kmps\ to $+7$~\kmps, 
which is red-shifted by 0$-$2.5~\kmps\ from the systemic velocity of the southern ridge ($\sim$$+4.5$~\kmps) 
and by 0.5$-$3~\kmps\ from that of the protostar (+4~\kmps). 
This strong localized emission can be confirmed in the velocity channel 
{\bff maps of $+5.4$ and $+7.1$~\kmps\ for both the \CO\ and SO emissions (Figures~\ref{fig:mom0-outflow}b, c, e, f).} 
The red-shifted feature is consistent with the previous reports for 
the eastern outflow lobe by \citet{Ceccarelli et al.(2002)} or the southeastern one by \citet{Bussmann et al.(2007)}. 
Among the spectra of the $^{12}$CO ($J=6-5$) reported by \citet{Ceccarelli et al.(2002)} (see their Figure~1), 
the spectrum at 7\arcsec\ south of the protostar is the one obtained at the position nearest to Position~A. 
The $^{12}$CO emission 
{\bff peaks at $+6$~\kmps\ and shows a wing there,} 
which was interpreted to trace the outflowing gas \citep{Ceccarelli et al.(2002)}. 
This $^{12}$CO feature, therefore, seems to correspond to 
the red-shifted component of the SO emission in our Figure~\ref{fig:SO-southernridge-PV}. 
Hence, the 
SO emission 
at Position~A is naturally interpreted as kinematic evidence of the interaction of the outflow with the southern ridge. 

Since Position~A is located within the north-south extension of the dust continuum emission {\bff (Figures~\ref{fig:continuum},~\ref{fig:ratio-Trot_SO}),} 
one may contemplate whether the warming is due to an additional protostar, perhaps with the larger southern ridge supplying mass. The distribution of the continuum emission around this local peak, however, is broad 
and not localized, in contrast to that toward the known protostar of Elias~29.
Furthermore, the SO emission does not show a clear velocity gradient, 
expected if gravity is playing a role \citep[e.g.][]{Oya et al.(2022)}. 
{\bff While the 2MASS Bands H and J data shows an intensity peak at the protostellar position of Elias 29, 
we do not find any peaks centered at Position A 
indicating a point-like young stellar object candidate 
by using Aladin Sky Atlas \citep{Bonnarel et al.(2000)}.} 
Although we cannot rule out the protostar solution completely, the evidence provides stronger support to  a shocked region.  In this case, the dust continuum emission has been enhanced due to both an increased density and temperature from the shock; 
such a feature was reported for 
a young low-mass protostellar source IRAS 16293$-$2422 by \citet{Maureira et al.(2022)}.

\subsection{Temperature Enhancements due to Jet Shock Interactions} \label{sec:jet}
{\bff The other clear hot spot, Position~E, 
is on the eastern side of the protostar. 
It is located inside the southeastern outflow cavity structure, 
{\bff whose central axis has the position angle (P.A.) of $\sim$120\degr,} 
seen in the \CO\ emission 
{\bff (Figure~\ref{fig:bowshock}a). 
Furthermore, 
it is nearly on the central axis of the parabolic shape of the outflow morphology 
represented by the pink sold line in Figure~\ref{fig:bowshock}(a), 
which likely corresponds to the outflow axis.} 
The emission at Position~E is evident in the velocity channel map of $+8.3$~\kmps\ as shown in Figure~\ref{fig:bowshock}(a). 
This emission is red-shifted by $+4.3$~\kmps\ with respect to the systemic velocity of the protostar ($\sim$$+4$~\kmps). 
{\bff This observed velocity-shift along the line-of-sight 
corresponds to a propagating velocity of 24.8~\kmps\ after correcting for projection 
using the inclination angle of 80\degr\ reported by \citet{Oya et al.(2019)}.} 
The direction {\bff from} 
this position 
{\bff to} 
the protostar {\bff (P.A.~285\degr)} 
roughly corresponds to the direction of the outflow cavity structure {\bff (P.A.~120\degr)} 
discussed in Section~\ref{sec:outflow}. 
Thus, one possible cause for this feature is a bow shock produced by the interaction 
between the protostellar jet and the ambient gas which has remained undissipated by the outflow. 

To verify this picture, we obtained PV diagrams around Position~E as shown in Figures~\ref{fig:bowshock}(b)~and~(c). 
The position axes are passing through Position~E (Figure~\ref{fig:bowshock}a).  
Figure~\ref{fig:bowshock}(b) shows the PV diagram obtained through Position~E and the protostellar position. 
The compact component concentrated to the protostar is seen 
as the bright emission with a wide velocity range from $-8$ to $+16$~\kmps\ at the angular offset of 0\arcsec. 
This disk/envelope component 
was previously analyzed by \citet{Oya et al.(2019)}. 
In addition to the compact feature, 
we see several bright knots toward the east of the protostar. 
The knot with velocity from $+8$ to $+10$~\kmps\ corresponds to the emission at Position~E. 
Taken together, these knots show a velocity gradient from the protostar to Position~E; 
they seem to be accelerated from the systemic velocity of $+4$~\kmps\ at the protostar. 
Figure~\ref{fig:bowshock}(c) shows the PV diagram across the outflow lobe passing through Position~E. 
In addition to the widely-extended ($\sim$20\arcsec) and slow (from $+2$ to $+6$~\kmps) component of the outflow structure, 
we find bright emission associated with a localized ($<$2\arcsec) and fast (from $+8$ to $+10$~\kmps) component. 
{\bff This behavior is naturally expected for a bow shock \citep[e.g.][]{Gustafsson et al.(2010)}; 
a bow shock presents the maximum acceleration of the gas at its apex, 
while the shocked gas is less perturbed away from the axis of the shock. 
Thus, the bright emission at Position~E is interpreted to be a bow shock produced by the narrow jet from the protostar. 
} 

{\bff 
It is worth noting that a feature morphologically similar to what we found in Figure~\ref{fig:bowshock}(b) 
was found in the SiO emission toward IRDC (infrared dark cloud) G034.77$-$00.55 \citep[Figure~4 by][]{Cosentino et al.(2019)} 
caused by an interaction with the supernova remnant W44, 
which was interpreted to be a magnetohydrodynamic C-shock. 
While the shock in G034.77$-$00.55 was suggested to be produced by the deceleration due to the dense material of the IRDC, 
the shocked gas in Elias~29 seems to accelerate. 
Thus, we need the shock modeling to examine whether the observed feature is related to a C-shock. 
} 

Assuming that the typical observed velocity of the bow shock ($+8$~\kmps; Figure~\ref{fig:bowshock}) 
is comparable to the propagating velocity projected onto the plane of the sky 
and that the systemic velocity is $+4$~\kmps, 
the dynamical time scale of the bow shock is roughly estimated to be 
\begin{equation}
t_{\rm dyn} = \frac{d_{\rm shock}}{v_{\rm jet}} \sim \frac{600}{\tan (i)}\ {\rm yr}. 
\end{equation}
Here, $d_{\rm shock}$ denotes the deprojected distance from the protostar to the bow shock, 
$v_{\rm jet}$ the deprojected propagating velocity of the jet, 
and $i$ the inclination angle of the jet axis with respect to the line of sight, 
where $i$ of 0\degr\ corresponds to the pole-on geometry. 
{\bff For instance, 
$t_{\rm dyn}$, $d_{\rm shock}$, and $v_{\rm jet}$ are obtained to be 
110~yr, 560~au, and 23~\kmps, respectively.} 
{\bff \citet{Oya et al.(2019)} reported the lower limit of 65\degr\ for $i$ 
based on the ellipticity of the continuum emission, 
and \citet{Oya et al.(2022)} reported that 
the rotation motion of the gas associated to the protostar 
is well reproduced by the Keplerian motion with $i$ of 80\degr. 
Then, we assume $i$ of 80\degr\ in the above calculation. 
If we employ the lower (65\degr) and upper limit (120\degr) for $i$ 
discussed in the above literatures, 
$t_{\rm dyn}$ is calculated to be 280~yr and 350~yr, respectively.} 
If this short time scale of the shock is the case, 
it would suggest that a mass accretion onto the protostar still occurs in this late Class I source. 
} 

\subsection{Mass Accretion in the Disk/Envelope System} \label{sec:accretion}

{\bff 
As described in Sections~\ref{sec:outflow} and \ref{sec:jet}, 
we found a young and energetic outflow/jet emerging from the protostar in this study. 
This result suggests that significant mass accretion is continuing at least sporadically. 
Although Elias~29 is a protostellar source at a late Class I stage, 
\citet{Oya et al.(2022)} reported that 
there is still a hint of infall motion of the gas in the compact structure in the vicinity of the protostar 
{\bff in addition to} 
the Keplerian motion. 

The relatively large bolometric luminosity \citep[$L_{\rm bol} = 13.6$~\Lsun;][]{Lommen et al.(2008), Miotello et al.(2014)} 
also suggests a moderate accretion rate. 
When we employ the protostellar mass ($M$) of 0.8~\Msun\ \citep{Oya et al.(2022)}, 
the mass accretion rate ($\dot{M}$) is roughly evaluated to be $1.4 \times 10^{-6}$ \Msun\ yr$^{-1}$ 
by using the following equation given by \citet{Palla & Stahler(1991)}: 
\begin{equation}
\dot{M} = \frac{L_{\rm bol} R}{G M}, \label{eq:Macc}
\end{equation}
where 
$G$ {\bff denotes} the gravitational constant. 
We here assume a protostar radius ($R$) 2.5 times the solar radius \citep{Larson(2003)}. 
This mass accretion rate is comparable to the typical value during the main accretion phase 
\citep[e.g.][]{Hartmann(2008)} 
and thus larger than that expected at the end of the Class I stage.
{\bff Such a high accretion rate ($>10^{-6}$ \Msun\ yr$^{-1}$) was also reported 
for {\bff a considerable number of} Class I sources 
by \citet{Enoch et al.(2009)} 
{\bff (Figure~12 in their paper). 
These authors concluded that the mass accretion during the Class I stage is episodic, 
considering the large dispersion in the observed bolometric luminosity.} 
} 

The disk mass derived from the continuum emission is reported to be 
$(7.3 \pm 0.6) \times 10^{-3}$ \Msun\ \citep{Artur de la Villarmois et al.(2019)}. 
Hence, 
the disk lifetime should be of {\bff the order of $10^{3-4}$ years,} 
if the accretion from the disk component toward the protostar is continuous at the rate evaluated above. 
This source does not have a massive surrounding envelope which could sustain a continuous accretion at the above rate, 
as revealed by the lack of such components in our \CO\ and CS data. 
With these considerations, 
the outflow/jet activities are expected to be sporadic rather than continuous, 
and the protostar would currently be in a relatively high accretion phase that cannot last for a long time. 

{\bff 
The indication of stronger accretion than expected for a late Class I source and the observation of a bow-shaped shock along the jet, 
within the southeast outflow lobe, 
suggests that Elias~29 should be investigated for potential accretion variability, 
as suggested for another luminous Class I source in Perseus by \citet{Valdivia-Mena et al.(2022)}. 
Non-steady mass assembly during the protostar stage is expected and observed, 
and can be an important probe of the underlying physical conditions in the disk \citep{Fischer23}.  
The JCMT (James Clerk Maxwell Telescope) 
Transient Survey \citep{Herczeg17} has been monitoring Ophiuchus, including Elias~29, 
for over eight years \citep{LeeYH21, Mairs24} at 850\,$\mu$m with a 15\arcsec\ beam and thus far has found no evidence of variability for this source. 
The same survey, however, has shown that at least 25\% of protostars are variable on years-long timescales, 
with these sources showing both episodic variations \citep[e.g. EC\,53;][]{LeeYH20} 
and burst behavior \citep[e.g. HOPS\,373;][]{Yoon22}. 
Therefore, continued monitoring of Elias~29 is strongly encouraged.
}

\section{Chemical Characteristics in the Shocked and Warm Environment} \label{sec:chem}
\subsection{{\bff Column Density} of SO at the Shock Locations} \label{sec:abundance}
We evaluated the 
column density of SO {\bff ($N({\rm SO})$)}
for the six positions shown in Figure~\ref{fig:ratio-Trot_SO} 
under the assumption of the LTE condition with \Trot\ derived in Section \ref{sec:res_temp} 
{\bff by using the following equation: 
\begin{equation}
N({\rm SO}) = U(T_{\rm rot}) \frac{3 k_{\rm B} W}{8 \pi^3 \nu\ S \mu^2} \exp \left(\frac{E_{\rm u}}{k_{\rm B} T_{\rm rot}} \right), 
\label{eq:nSO}
\end{equation} 
where $U(T_{\rm rot})$ denotes the partition function of SO 
at the temperature $T_{\rm rot}$, 
and $k_{\rm B}$ the Boltzmann constant.
The line parameters (the frequency $\nu$, $S \mu^2$, and $E_{\rm u}$) 
are summarized in Table~\ref{tb:molecule}, 
while the integrated intensities are in Table~\ref{tb:SOparams}.} 
{\bff Since the integrated intensity $W$ is used, 
the derived column density is an averaged one over velocity components 
along the line of sight.} 
The results of the LTE analysis are summarized in Table~\ref{tb:temp-abundance_SO}. 
{\bff The obtained column density of SO 
is similar to each other among Positions~B$-$E 
($(1.0-1.7) \times 10^{14}$~\cminvsq) 
while it is about 5 times higher at Position~A ($8.4 \times 10^{14}$~\cminvsq). 
Only the lower limit was obtained at the continuum peak position 
{\bff due to the high opacity of the SO lines.}} 
We also conducted a non-LTE analysis for confirmation in Appendix \ref{sec:appendix} 
and found that the column densities derived under the LTE assumption 
{\bff are almost consistent with those derived by the non-LTE analysis, 
except for {\bff Position~D} with a large error; 
the difference of the evaluated values are within 20\%\ for Positions~A, B, and E between the two analyses, 
while the evaluated values are different by a factor of 2 for Position~C.} 
The error for the non-LTE result is large 
because 
the \hydro\ density {\bff is treated as a free parameter} 
as well as the gas kinetic temperature and the column density 
{\bff (see Appendix~\ref{sec:appendix}).} 
} 

{\bff 
In principle, 
it is desirable to derive the SO abundance relative to \hydro\ by 
using the \hydro\ estimated from the \CO\ data. 
However, 
the line profiles of \CO\ at \posAE\ are 
different from those of SO (Figure~\ref{fig:spectra-SO}). 
The \CO\ line has a lower critical density than the SO lines 
and would preferentially trace the different parts along the line of sight. 
Thus, it is almost impossible to disentangle these components 
and derive their SO abundances reasonably. 
Hence, we just discuss the SO column densities in this paper. 
}

SO is often recognized as a shock tracer \citep[e.g.][]{Bachiller & Perez Gutierrez(1997)}; 
its abundance in the gas phase is expected to be {\bff increased} 
by the sputtering of S atoms and/or SO from dust or ice mantle in shocked regions. 
{\bff The SO column density at the outflow interaction region (Position~A) 
is indeed higher than those at other positions in the southern ridge (\posBD), 
which would indicate a sign of shock enhancement. 
However, 
the SiO (\sioa) line is undetected toward Position~A in our observation,} 
although 
{\bff the} SiO emission has been detected and associated with jets and strong shocks in other sources 
\citep[e.g.][]{Mikami et al.(1992), Bachiller & Perez Gutierrez(1997), Zapata et al.(2009), Feng et al.(2016), Oya et al.(2018), Oya et al.(2018) 16293B, Okoda et al.(2021), Sato et al.(2023)} 
due to the destruction and sputtering of silicate dust grains 
\citep{Ziurys et al.(1989), Caselli et al.(1997), Schilke et al.(1997)}. 
{\bff Therefore, 
the shock at Position~A} 
may not be enough strong to destroy or sputter dust grains 
to liberate SiO into the gas phase. 
The high column density of SO at Position~A 
{\bff may alternatively be} 
due to the high gas column density at that location. 
{\bff In this case, the SO molecules} 
are not originating from {\bff dust/ice} mantle, 
but are already in the gas-phase prior to the shock. 

For Position~E, 
{\bff the SO column density is comparable to the quiescent parts 
of the southern ridge (\posBD). 
Moreover,} 
a lack of SiO emission {\bff at Position~E} is puzzling, 
{\bff considering that the bow shock feature is observed.} 
A detailed relation of the Position~E to the outflow is not fully understood 
and might need further 
{\bff investigation both in observations and modeling.}

\citet{Oya et al.(2019)} previously suggested that 
the warm environment of Elias~29 prevents C atoms and CO molecules from being adsorbed onto dust grains 
in the envelope such that this source is deficient in organic molecules. 
If this is also the case for S atoms, 
the majority of SO molecules would need to be formed and preserved in the gas-phase. 
{\bff This is likely the case 
because the desorption temperature of S atoms (1100~K) 
is less than or comparable to that of C atoms (800~K) and CO molecules (1150~K) 
according to Kinetic Database for Astrochemist 
\citep[KIDA;][{\url http://kida.obs.u-bordeaux1.fr/}]{Wakelam et al.(2012)}, 
as discussed by \citet{Oya et al.(2019)}.} 

\subsection{Deuterium Fractionation Ratio in the Southern Ridge} \label{sec:warm_dco}

The line width of the SO emission is less than 1~\kmps\ at \posBD\ as shown 
in Table~\ref{tb:SOparams} and {\bff Figures~\ref{fig:spectra-SO} and \ref{fig:SO-southernridge-PV}.} 
Hence, 
it is difficult to explain the moderately high rotational temperature of SO (20$-$30~K; Table~\ref{tb:temp-abundance_SO}) at these three positions
in terms of the {\bff recent} 
outflow shock. 
This situation is different from that at Positions~A and E with larger line width, 
where interactions {\bff with} 
the outflow/jet are seen in the velocity structure as discussed in Sections \ref{sec:outflow} and \ref{sec:jet}. 
Thus, it is important to assess these derived moderately high temperatures in an independent way. 

{\bff As an indicator for the temperature environment, 
we investigate the deuterium fractionation ratio of \HCOmain\ in the southern ridge component. 
For this purpose, 
we use the \HCO\ (\hcoa) and \DCO\ (\dcoa) emission, 
assuming the constant [\HCOmain]/[\HCO] ratio of 60 \citep{Lucas & Liszt(1998)}.} 
The spectra {\bff of \HCO\ and \DCO\ emission} 
toward \posAD\ are shown in Figure~\ref{fig:H13COp-DCOp-spectra}, 
and {\bff the line parameters are} summarized in Table~\ref{tb:hco-dco-params}. 
{\bff The observed peak intensity of the \DCO\ spectrum 
is lower than the noise level 
at Position~D, 
and thus the 2$\sigma$ noise error is employed as the upper limit.} 
We employ a non-LTE analysis for both the two molecular lines, 
as we performed for the SO lines (see Appendix \ref{sec:appendix} for the details). 
\hydro\ is assumed to be the collision 
partner. 
Since the collisional cross sections of \HCO\ are not available, 
we employ those of \HCOmain\ for \HCO. 
The collisional cross sections of \HCOmain\ and \DCO\ are taken from} 
{\bff Leiden Atomic and Molecular Database \citep[LAMDA;][]{Schoier et al.(2005)},} 
whose original data are reported by \citet{Denis-Alpizar et al.(2020)}. 
{\bff For \HCOmain, 
the collisional cross sections with para-\hydro\ and ortho-\hydro\ are separately listed. 
We assume the ortho-para ratio of \hydro\ to be 3 in the analysis. 
Note that the following results do not change significantly by using an ortho-to-para ratio of 0.} 

In the \chisq\ analysis, 
{\bff we use the peak intensities of the \HCO\ and \DCO\ intensities listed in Table~\ref{tb:hco-dco-params}. 
The} 
\chisq\ values for \HCO\ and \DCO\ were summed up to obtain the total \chisq\ value, 
which is used to constrain the ranges of the \HCO\ column density and the [\DCO]/[\HCO] ratio. 
In this analysis, 
the ranges of the gas kinetic temperature and the \hydro\ number density are assumed to be 
those derived in the non-LTE analysis for the SO emission 
(Tables~\ref{tb:temp-abundance_SO}). 

The {\bff estimated} [\DCO]/[\HCOmain] ratio is summarized in Table~\ref{tb:DFR}. 
The [\DCO]/[\HCOmain] ratio is below {\bff 1.5\%} including the error for all the positions. 
These values are relatively low in comparison with those found in typical cold prestellar cores \citep[$\sim$4\%;][]{Bacmann et al.(2003)}. 
This result 
indicates a relatively warm condition for the entire southern ridge. 
This is consistent with the result for the rotational temperature and the gas kinetic temperature derived for SO (Section \ref{sec:res_temp}). 

It is interesting that a moderately high temperature is found even in the quiescent part of the southern ridge. 
One possible cause for this result is the external illumination by \Bstar, a B-type star at 700\arcsec\ ($\sim$0.5 pc) northwest of Elias~29. 
This star is suggested to heat the entire area, 
and even the whole Ophiuchus molecular cloud. 
Indeed, \citet{Rocha & Pilling(2018)} reported 
that the molecular cloud core of Elias~29 is warmed up to more than 20 K by \Bstar\ {\bff based on their radiative transfer modeling.} 

The relatively warm conditions in the southwestern part of the southern ridge 
without apparent influence of the on-going outflow interaction 
may also explain the lack of iCOMs and hydrocarbon molecules in Elias~29. 
If the gas temperature of the parent cloud of Elias~29 were higher than 20~K due to the irradiation from \Bstar, 
CO molecules and C atoms would not deplete onto dust grains in the prestellar core phase. 
As a result, iCOMs and CH$_4$ are not produced efficiently via dust surface reactions \citep{Oya(2022)}. 
This possibility was pointed out by \citet{Oya et al.(2019)}, 
and it is further strengthened by the temperature structure 
revealed in this project. 
Elias~29 is a novel example 
where the combination of environmental 
dynamics and radiation triggers peculiar chemistry. 
This situation is possibly similar to that of 
{\bff the photodissociation region (PDR)} 
R~CrA~IRS7B \citep{Watanabe et al.(2012), Lindberg et al.(2015)} 
where the abundance of iCOMs has been found to be unexpectedly deficient in a single dish survey. 
On the other hand, 
the difference is the abundance of the CCH emission, 
{\bff which is known to be enhanced by the PDR (photodissociation region).  
The fractional abundance of CCH to \hydro\ is reported to be 
$(5.3\pm1.5) \times 10^{-9}$ in R~CrA~IRS7B with the gas temperature assumed to be 20~K \citep{Watanabe et al.(2012)}, 
while its upper limit is $(1.3-9.0) \times 10^{-11}$ in Elias~29 with the gas and dust temperatures assumed to be from 50 to 150~K \citep{Oya et al.(2019)}.}  
{\bff This large difference in the fractional abundance of CCH by the two order of magnitude} 
implies that the PDR 
chemistry is not working in Elias~29 
probably due to heavier attenuation of the far UV photons dissociating CO.

\section{Summary} \label{sec:summary}
In order to characterize the protostellar activity of the late Class I source Elias~29, we observed lines of SO, \SO, \CO, CS, SiO, \HCO, and \DCO\   
at a spatial resolution of 70~au (0\farcs5). 
This investigation is part of the ALMA large program FAUST. 
Our major findings are summarized below. 

\begin{itemize}
\item[(1)]
We delineated the temperature distribution by using 
{\bff two transitions of SO and one of \SO.} 
We obtained a two-dimensional map of the rotational temperature of SO, using LTE assumptions, 
and found two hot spots 
{\bff with the rotational temperature of $53^{+25}_{-15}$ and $72^{+66}_{-29}$~K} 
in addition to the hot component associated with the protostar. 
These locally high temperatures were confirmed with a non-LTE analysis. 
\item[(2)] 
The local increase of the temperature 
{\bff at the northeastern tip of} 
the southern ridge component 
is likely attributed to interaction with the southeast outflow lobe. 
The other local temperature increase 
is interpreted as the result of a bow shock produced by the jet. 
It is interesting that 
Elias~29 still has significant outflow/jet activity {\bff at least sporadically} 
in spite of its classification as a late-type Class I. 
\item[(3)] 
It is likely that significant mass accretion from disk to protostar is still occurring at least sporadically, 
considering the energetic activity of the outflow/jet in Elias~29 observed in this project, 
{\bff the hint of the infall motion in its disk/envelope system previously suggested, 
and its relatively large luminosity.} 
{\bff The mass accretion rate is roughly evaluated 
to be $1.4 \times 10^{-6}$ \Msun\ yr$^{-1}$, 
which sounds high for the late Class I stage of this source.} 
Since the outer envelope has already been consumed/dissipated, 
as demonstrated by the lack of the \CO\ emission associated around the protostar, 
we are likely witnessing the near end of the mass accretion phase. 
\item[(4)]
{\bff We evaluated the SO column density 
at the five positions (\posAE). 
Although the SO column density 
at the outflow-ridge interaction region (Position~A) 
is found to be higher by a factor of 5 
than in the quiescent part of the southern ridge (\posBD), 
the SiO emission is not detected there.} 
{\bff {\bff This result} 
{\bff may indicate} 
that the shock is not strong enough to destruct or sputter the silicate dust grains {\bff at Position~A.}}
{\bff In spite of the bow-shock feature at Position~E, 
the SiO emission is not detected there either, 
and the SO column density is comparable to those at \posBD.} 
{\bff {\bff The} 
origin of {\bff these results} 
at Position~E is still puzzling; 
further observational and modeling efforts are necessary.} 
\item[(5)] 
Even the quiescent parts of the southern ridge at distances ranging from 500 to 1000 au away from the protostar are found to be moderately warm (20$-$30 K). 
This warm condition is consistent with our {\bff estimated} relatively low deuterium fractionation (1\% or less) in \HCOmain. 
The B-type star \Bstar, which is 700\arcsec\ away from Elias~29, 
would likely have kept the parent cloud of Elias~29 warm. 
Such a warm environment for {\bff the} entire lifetime of this source could be the cause for its peculiar chemical characteristics 
{\bff including a deficiency of} 
iCOMs and hydrocarbon molecules. 
\end{itemize}

\acknowledgments
This paper makes use of the following ALMA data: ADS/JAO.2018.1.01205.L. ALMA is a partnership of ESO (representing its member states), NSF (USA) and NINS (Japan), together with NRC (Canada), NSC and ASIAA (Taiwan), and KASI (Republic of Korea), in cooperation with the Republic of Chile. 
The Joint ALMA Observatory is  operated by ESO, AUI/NRAO and NAOJ. 
This study is supported by Grant-in-Aids from Ministry of Education, Culture, Sports, Science, and Technologies of Japan (18H05222, 18J11010, 19H05069, 19K14753, and 21K13954). 
The authors acknowledge the financial support by JSPS and MAEE under the Japan-France integrated action programme (SAKURA: 25765VC). 
{\bff Y.O. acknowledges NAOJ for the ALMA Joint Scientific Research Program (2024-27B).} 
Y.O. {\bff also thanks} NAOJ for the support as the ALMA 10th Anniversary Award. 
D.J.\ is supported by NRC Canada and by an NSERC Discovery Grant.
LP and CC acknowledge financial support under the National Recovery and Resilience Plan (NRRP), Mission 4, Component 2, Investment 1.1, Call for tender No. 104 published on 2.2.2022 by the Italian Ministry of University and Research (MUR), funded by the European Union - NextGenerationEU - Project Title 2022JC2Y93 ChemicalOrigins: linking the fossil composition of the Solar System with the chemistry of protoplanetary disks - CUP J53D23001600006 - Grant Assignment Decree No. 962 adopted on 30.06.2023 by the Italian Ministry of Ministry of University and Research (MUR).
LP, CC, and GS also acknowledge the EC H2020 project “Astro-Chemical Origins” (ACO, No 811312), the PRIN-MUR
2020 BEYOND-2p (Astrochemistry beyond the second period elements, Prot. 2020AFB3FX), the project ASI-Astrobiologia 2023
MIGLIORA (Modeling Chemical Complexity, F83C23000800005), the INAF-GO 2023 fundings
PROTO-SKA (Exploiting ALMA data to study planet forming disks: preparing the advent of SKA,
C13C23000770005), the INAF-Minigrant 2022 “Chemical Origins” (P.I.: L. Podio), and the INAF-Minigrant 2023 TRIESTE (“TRacing the chemIcal hEritage of our originS: from proTostars to plan- Ets”; PI: G. Sabatini).
L.L. acknowledges the support of UNAM-DGAPA PAPIIT grants IN112820 and IN108324 and CONAHCYT-CF grant 263356.
This project has received funding from the European Research Council (ERC) under the European Union Horizon Europe research and innovation program (grant agreement No. 101042275, project Stellar-MADE). 

S.V. and M.B. acknowledge support from the ERC  under the European Union's Horizon 2020 Research and innovation programm MOPPEX 833460.
G.S. also acknowledges support from the INAF-Minigrant 2023 TRIESTE (``TRacing the chemIcal hEritage of our originS: from proTostars to planEts''; P.I.: G. Sabatini). 
I.J-.S acknowledges funding from grants No. PID2019-105552RB-C41 and PID2022-136814NB-I00 from the Spanish Ministry of Science and Innovation/State Agency of Research MCIN/AEI/10.13039/501100011033 and by ``ERDF A way of making Europe''.
E.B. acknowledges the Deutsche Forschungsgemeinschaft (DFG, German Research Foundation) under Germany's Excellence Strategy -- EXC 2094 -- 390783311.
S.B.C was supported by the NASA Planetary Science Division Internal Scientist Funding Program through the Fundamental Laboratory Research work package (FLaRe).
{\bff E.B. acknowledges contribution of the Next Generation EU funds within the National Recovery and Resilience Plan (PNRR), Mission 4 - Education and Research, Component 2 - From Research to Business (M4C2), Investment Line 3.1 - Strengthening and creation of Research Infrastructures, Project IR0000034 - “STILES - Strengthening the Italian Leadership in ELT and SKA.} 

\appendix 
\section{Non-LTE Analysis for SO} \label{sec:appendix}
To confirm whether the results obtained by the LTE (local thermodynamic equilibrium) analysis for SO 
(Figure~\ref{fig:ratio-Trot_SO}b; Section \ref{sec:res_temp}) is reasonable, 
we conduct a non-LTE analysis. 
The non-LTE analysis is performed for five picked-up positions (A through E), 
unlike the pixel-based method for the calculation of the rotational temperature of SO. 

We use the \SO\ (\thsoa) line data in the non-LTE analysis in addition to the two SO (\soa, \sob) lines 
for accurate determination of the gas kinetic temperature. 
Figure~\ref{fig:spectra-SO} shows the spectra of these three lines at \posAE. 
Although the \SO\ emission is much fainter than the SO emission, 
{\bff it is detected toward \posAD\ with the S/N ratio higher than 6.} 
Meanwhile, the detection of the \SO\ emission is marginal at Position~E, 
{\bff where the S/N ratio is 3.6.} 
The parameters of the spectra are summarized in Table~\ref{tb:SOparams}. 
The \SO\ line is frequency-diluted, 
since it is observed with a coarse-resolution backend (1.129 MHz). 
Hence, its peak intensity at each position is derived from the integrated intensity 
divided by the averaged line width of the two SO lines. 

We conduct the 
non-LTE 
modeling and constrain the gas kinetic temperature, the SO column density, 
and the \hydro\ number density by the \chisq\ method. 
{\bff We employed the escape probability for 
a static, spherically symmetric, and homogeneous medium 
\citep{Osterbrock & Ferland(2006), van der Tak et al.(2007)}.} 
We employ the energy levels, the Einstein A coefficients, and the state-to-state collisional rates of SO 
by \citet{Price et al.(2021)}. 
The energy levels and the Einstein A coefficients of \SO\ are taken from CDMS 
{\bff \citep{Muller et al.(2005), Endres et al.(2016)}.} 
{\bff We employ \hydro\ as the collision 
partner\footnote{The ortho-to-para ratio of \hydro\ is assumed to be 3 in this calculation. 
We confirmed that the results do not change significantly by using an ortho-to-para ratio of 0.}.} 
Since the collisional rates of \SO\ are not available, 
we employ those of SO as substitutes without any corrections. 
We assume the ratio of $N \left(^{32}{\rm SO}\right) / N \left(^{34}{\rm SO}\right)$ to be 19 in this calculation 
\citep{Lucas & Liszt(1998)}. 

In the \chisq\ analysis, 
{\bff the SO and \SO\ data are treated simultaneously in the \chisq\ analysis. 
Thus, the \chisq\ value is calculated by the following equation; 
\begin{equation}
\chi^2 = \Sigma_i \left( \frac{T_i^{\rm obs} - T_i^{\rm calc}}{\sigma_i^2} \right)^2, 
\label{eq:chisq}
\end{equation}
where the summation is taken over the two SO and one \SO\ lines. 
$T_i^{\rm obs}$ and $T_i^{\rm calc}$ denote 
the observed and calculated peak temperatures. 
$\sigma_i$ stands for the error of the observed temperature 
which is calculated from the two times the root-mean-square (rms) noise level 
of each spectrum and 10\%\ of the intensity calibration error. 
$T_i^{\rm obs}$ and its rms are summarized in Table~\ref{tb:SOparams}. 

In the analysis,} 
we scan wide ranges of parameter values to find the best-fit parameter values: 
every 2.5~K from 30 to 117.5~K for the gas kinetic temperature, 
every $0.25 \times 10^{14}$~\cminvsq\ from 8 to $13.75 \times 10^{14}$~\cminvsq\ for the SO column density, 
and every 1.5 times from $3 \times 10^5$ to $5.05 \times 10^9$~\cminvcb\ for the \hydro\ number density. 
The results are summarized in Table~\ref{tb:temp-abundance_SO}. 
The error values denote 
the range of the parameter value 
{\bff which corresponds to the above intensity error.} 

At Position~A, the interaction position of the outflow and the southern ridge component, 
the gas kinetic temperature is derived to be $55^{+39}_{-14}$ K. 
The relatively high temperature derived for Position~A indicates 
that the shock heating is indeed occurring there, as discussed in Section \ref{sec:outflow}. 
The gas kinetic temperatures are evaluated to be $>23$ K, $26^{+15}_{-3}$, and $>16$ K 
at Positions~B, C, and D in the southern ridge, respectively. 
Although only the lower limits are obtained for Positions~B and D, 
the gas kinetic temperature tends to decrease as leaving from Position~A, and also from the protostar, along the southern ridge; 
however, it is still higher than 16 K at Position~D, which is the end of the southern ridge. 

Only the lower limit is available for the gas kinetic temperature at Position~E. 
Since the \SO\ emission is marginally seen, 
the constraint on the parameters is rather loose (Table~\ref{tb:temp-abundance_SO}). 
Nevertheless, the high temperature ($>$56 K) condition is indeed confirmed, 
as suggested by the rotational temperature of SO (Figure~\ref{fig:ratio-Trot_SO}). 
This result supports the local heating occurring at Position~E, as well as the result for Position~A; 
the heating mechanism is likely attributed to the bow shock at Position~E, as discussed in Section \ref{sec:jet}. 

\ifshortauthorslist
\allauthors
\fi

{}


\clearpage
\scalebox{0.8}{
\centering
\begin{threeparttable}
\caption{List of the Observed Lines$^{\rm a}$}
\label{tb:molecule}
\centering
\small 
\begin{tabular}{cccccc}
\hline  
Molecule & Frequency (GHz) & $S\mu^2$ (D$^2$) &  $E_{\rm u}$ (K) & Beam size$^{\rm b}$ & rms (mJy beam$^{-1}$)\\ \hline 
\co\ (\coa) 	& 219.5603541	& 0.0244	& 16	& $0\farcs531 \times 0\farcs438$ (P.A. $-89.\!\!\degr463$)	& 1.8\\			
SO (\soa) 		& 219.949442 	& 14.0	& 35	& $0\farcs528 \times 0\farcs435$ (P.A. $89.\!\!\degr739$)	& 2.3$^{\rm c}$ \\ 	
SO (\sob) 		& 258.2558259	& 13.7	& 57	& $0\farcs573 \times 0\farcs451$ (P.A. $-79.\!\!\degr952$)	& 2.5$^{\rm c}$ \\ 	
\SO\ (\thsoa) 	& 246.66347	& 11.4	& 50	& $0\farcs518 \times 0\farcs392$ (P.A. $-73.\!\!\degr648$)	& 1.3\\ 			
CS (\csa) 		& 244.9355565	& 19.2	& 35	& $0\farcs523 \times 0\farcs402$ (P.A. $-73.\!\!\degr669$)	& 1.8\\ 			
SiO (\sioa) 	& 217.1049190	& 48.0	& 31	& $0\farcs537 \times 0\farcs439$ (P.A. $-88.\!\!\degr865$)	& 1.7 \\ 			
\HCO\ (\hcoa) 	& 260.255339	& 45.6	& 25	& $0\farcs497 \times 0\farcs381$ (P.A. $-73.\!\!\degr099$)	&  2.8\\ 			
\DCO\ (\dcoa) 	& 216.1125822	& 45.6	& 21	& $0\farcs540 \times 0\farcs442$ (P.A. $-88.\!\!\degr636$)	&  2.3\\ 			
\hline
\end{tabular}
\begin{tablenotes}
\item[$^{\rm a}$] Taken from CDMS \citep{Muller et al.(2005), Endres et al.(2016)}. 
\item[$^{\rm b}$] {\bff Synthesized beam.} 
\item[$^{\rm c}$] 
{\bff Root-mean-square noise} 
of the data with the beam size of $0\farcs6 \times 0\farcs6$. 
\end{tablenotes}
\end{threeparttable}
}

\clearpage
\scalebox{0.7}{
\begin{threeparttable}
\caption{Observation Parameters}
\label{tb:ObP}
\begin{tabular}{cccccccccc} 
\hline  
& Configuration & Date & $\rm{T_{\rm target}}$ &  $\rm{N_{Antennas}}$ & Baseline & MRS$^{\rm a}$ & \multicolumn{3}{c}{Calibrator}  \\ 
& & & (min) & & (meter) & (arcsec) & Phase & Bandpass & Flux\\ \hline
\multirow{6}{*}{Setup 1} 
& C43-4 & 2018 Dec 2	& 29.35 & 43 & 15.1$-$783.5 	& 5.7 	& J1625$-$2527 & J1427$-$4206 & J1427$-$4206\\  
& C43-4 & 2019 May 4	& 29.35 & 40 & 15.1$-$740.4 	& 5.5 	& J1626$-$2951 & J1427$-$4206 & J1427$-$4206\\  
& C43-4 & 2020 Mar 16	& 29.33 & 44 & 15.2$-$968.7 	& 5.4 	& J1626$-$2951 & J1517$-$2422 & J1517$-$2422 \\ 
& C43-1 & 2019 Jan 21	& 10.12 & 48 & 15.0$-$313.7 	& 11.3 	& J1625$-$2527 & J1517$-$2422 & J1517$-$2422 \\ 
& ACA    & 2018 Nov 18	& 34.83 & 10 & 8.9$-$44.7   	& 30.3	& J1625$-$2527 & J1256$-$0547 & J1256$-$0547\\  
& ACA    & 2018 Nov 22	& 34.78 & 12 & 8.9$-$48.9   	& 30.3	& J1625$-$2527 & J1337$-$1257 & J1337$-$1257\\  
\hline
\multirow{5}{*}{Setup 2} 
& C43-4  & 2018 Dec 1	& 20.73 & 43 & 15.1$-$951.7 	& 5.3 	& J1625$-$2527 & J1427$-$4206 & J1427$-$4206\\  
& C43-4  & 2019 Apr 17	& 20.75 & 46 & 15.1$-$740.4 	& 5.6 	& J1650$-$2943 & J1517$-$2422 & J1517$-$2422\\  
& C43-4  & 2019 Oct 11	& 20.72 & 43 & 15.1$-$783.5	& 5.5 	& J1626$-$2951 & J1427$-$4206 & J1427$-$4206\\  
& C43-1  & 2019 Jan 21	& 7.10   & 48 & 15.0$-$313.7 	& 10.0	& J1625$-$2527 & J1517$-$2422 & J1517$-$2422\\ 
& ACA     & 2018 Oct 25	& 49.43 & 11 & 8.9$-$48.9   	& 26.9	& J1700$-$2610 & J1924$-$2914 & J1924$-$2914\\ 
\hline
\end{tabular}
\begin{tablenotes}
\item[$^{\rm a}$] Maximum recoverable scale. 
\end{tablenotes}
\end{threeparttable}
}

\clearpage
\scalebox{0.9}{
\centering
\begin{threeparttable}
\caption{Coordinates of the Positions~A-E and C.P.} 
\label{tb:coords}
\begin{tabular}{cccc}
\hline 
Position~& Right Ascension (ICRS) & Declination (ICRS) & Note 
\\ \hline 
A & $16^{\rm h}27^{\rm m}09.\!\!^{\rm s}476$ & $-24\degr37\arcmin23\farcs533$ & Outflow interaction in the southern ridge \\ 
B & $16^{\rm h}27^{\rm m}09.\!\!^{\rm s}251$ & $-24\degr37\arcmin26\farcs104$ & 4\arcsec\ from Position~A along the southern ridge \\ 
C & $16^{\rm h}27^{\rm m}09.\!\!^{\rm s}139$ & $-24\degr37\arcmin27\farcs390$ & 6\arcsec\ from Position~A along the southern ridge \\ 
D & $16^{\rm h}27^{\rm m}09.\!\!^{\rm s}027$ & $-24\degr37\arcmin28\farcs675$ & 8\arcsec\ from Position~A along the southern ridge\\ 
E & $16^{\rm h}27^{\rm m}09.\!\!^{\rm s}692$ & $-24\degr37\arcmin20\farcs043$ & Bow shock by the jet\\ 
C.P. & $16^{\rm h}27^{\rm m}09.\!\!^{\rm s}41737$ & $-24\degr37\arcmin19\farcs3104$ & 1.2 mm continuum peak position \\ 
\hline 
\end{tabular}
\begin{tablenotes}
\item 
{\bff Spectra of the molecular lines are obtained 
as the averages over the circular region 
with a diameter of 1\arcsec\ centered at each of \posAD. 
Meanwhile, 
those with the original synthesized beam is employed 
for Positions~E and C.P.}
\end{tablenotes}
\end{threeparttable}
} 

\clearpage
\scalebox{0.9}{
\centering
\begin{threeparttable}
\caption{Parameters of the SO (\soa, \sob) and \SO\ (\thsoa) Emissions} 
\label{tb:SOparams}
\centering
\small 
\begin{tabular}{cccccc} \hline  
Position 
& Line & $T_{\rm peak}$/K $^{\rm a}$ & $V_{\rm LSR}$/ \kms \ $^{\rm b}$  &  $\Delta v$/ \kms \ $^{\rm c}$ & $W$/K km s$^{-1}$ $^{\rm d}$ \\  \hline
\multirow{3}{*}{A} 
& SO (\soa)  & 24.3(24) & 5.24  & 2.32  & 56.6(57) \\
& SO (\sob) & 20.2(20)   & 5.02  & 2.15  & 43.4(43) \\  
& \SO\ (\thsoa) & 1.19(13) & $-$ &  $-$    & {\bff $2.67(13)$}         \\  \hline
\multirow{3}{*}{B} 
& SO (\soa)  & 16.57(17)    & 4.56  & 0.60  & 11.3(12) \\  
& SO (\sob) & 9.92(99)  & 4.56  & 0.53  & 6.1(7)  \\  
& \SO\ (\thsoa) & 0.48(58)    & $-$ &  $-$    & {\bff $0.27(16)$}        \\ \hline
\multirow{3}{*}{C} 
& SO (\soa) & 18.5(19)   & 4.42  & 0.64  & 12.9(14) \\  
& SO (\sob) & 13.5(14)   & 4.33  & 0.53  & 7.9(9)  \\  
& \SO\ (\thsoa) & 1.26(59) & $-$ &  $-$    & {\bff $0.73(17)$}        \\  \hline
\multirow{3}{*}{D} 
& SO (\soa) & 12.0(12)   & 4.56  & 0.76  & 10.4(11) \\  
& SO (\sob) & 6.53(65)    & 4.67  & 0.62  & 4.5(7)  \\  
& \SO\ (\thsoa)  &  0.83(65)  & $-$ &  $-$    & {\bff $0.6(2)$}       \\  \hline
\multirow{3}{*}{E} 
& SO (\soa) & 3.8(5)   & 8.50(8) & 1.79(18) & 6.1(7) \\  
& SO (\sob) & 3.3(4)    & 8.42(6)  & 1.70(14)  & 5.2(6)  \\  
& \SO\ (\thsoa)  &  0.26(14)  & 10.0(5) & 1.75(fixed)  & 0.4(2)  \\  \hline
\multirow{3}{*}{C.P.} 
& SO (\soa) & $-$  & $-$ & $-$ & 201(20) \\  
& SO (\sob) & $-$  & $-$ & $-$  & 230(23)  \\  
& \SO\ (\thsoa)  &  $-$  & $-$  & $-$  & 21(3) \\  
\hline
\end{tabular}
\begin{tablenotes}
\item 
The numbers in parentheses represent the error in units of the last significant digits. 
\item[$^{\rm a}$] Peak intensity 
{\bff of the observed line profiles for \posAD.} 
{\bff The error is evaluated from twice the rms noise of each spectrum 
and the 10\%\ intensity calibration error, 
where the latter contribution dominates the total error.} 
{\bff Since the \SO\ emission is frequency-diluted due to a coarse-resolution backend, 
its peak intensity is derived from the integrated intensity 
divided by the averaged line width of the two SO lines (see Appendix~\ref{sec:appendix}).
For Position~E, the peak intensity is derived by the Gaussian fitting.} 
\item[$^{\rm b}$] 
{\bff LSR velocity of the intensity peak of the observed line profiles for \posAD.} 
For Position~E, the Gaussian fitted value is employed. 
\item[$^{\rm c}$] FWHM of the line width. 
The Gaussian fitted value is employed. 
Although the spectrum is not well fitted to define the peak intensity due to the asymmetric line shape for \posAD, 
the line width obtained by the fit is approximately employed as the best effort. 
\item[$^{\rm d}$] {\bff Observed} integrated intensity from 0.0 to $+10.0$~\kmps\ {\bff for \posAD, 
{\bff that from $+7.0$ to $+10.5$~\kmps\ for Position~E,} 
and that from $-15.0$ to $+23.0$~\kmps\ for the continuum peak (C.P.) position.} 
\end{tablenotes}
\end{threeparttable}
}

\clearpage
\scalebox{0.9}{
\centering
\begin{threeparttable}
\caption{Column Density and Rotational Temperature of SO at Representative Positions} 
\label{tb:temp-abundance_SO}
\begin{tabular}{ccccccc}
\hline 
& \multicolumn{2}{c}{LTE$^{\rm a}$} & \quad & \multicolumn{3}{c}{non-LTE$^{\rm b}$} \\ 
Position 
& $N({\rm SO}) / 10^{14}$ \cminvsq & \Trot\ (K) & & $N({\rm SO}) / 10^{14} {\rm cm}^{-2}$ & \Tkin\ (K) & $n({\rm H}_2) / 10^6$ \cminvcb \\ \hline 
A & $8.4^{+1.2}_{-0.6}$ & $53^{+25}_{-15}$ & 	& $10.7^{+1.2}_{-1.7}$ & $55^{+39}_{-14}$ & $>1.2$ \\ 
B & $1.5^{+0.2}_{-0.2}$ & $28^{+7}_{-5}$ & 	& $1.3^{+2.6}_{-0.2}$ & $>23$ & $>0.2$ \\ 
C & $1.7^{+0.2}_{-0.1}$ & $34^{+10}_{-7}$ & 	& $3.6^{+3.1}_{-1.6}$ & $26^{+15}_{-3}$ & $>0.4$ \\ 
D & $1.5^{+0.3}_{-0.2}$ & $22^{+5}_{-4}$ & 	& $7.3^{+11.0}_{-6.3}$ & $>16$ & $>0.11$ \\ 
E & $1.0^{+0.5}_{-0.2}$ & $72^{+66}_{-29}$ & 	& $1.2^{+0.8}_{-0.4}$ & $>56$ & $>0.4$ \\ 
C.P. & $>71$ & $>192$ & 					& $-$ & $-$ & $-$ \\ 
\hline 
\end{tabular}
\begin{tablenotes}
\footnotesize
\item[$^{\rm a}$] 
Values are derived under the assumptions of the LTE and optically thin conditions. See Sections~\ref{sec:res_temp} and \ref{sec:abundance}. 
{\bff The observed integrated intensities are summarized in Table~\ref{tb:SOparams}.} 
The error is derived from the 2$\sigma$ noise error in the integrated intensities and the intensity calibration error of 10\%. 
\item[$^{\rm b}$] Values are derived by the non-LTE analyses. See Appendix \ref{sec:appendix} for the details.  
{\bff The observed peak intensities are summarized in Table~\ref{tb:SOparams}.} 
The error is derived from the 2$\sigma$ noise error in the intensities and the intensity calibration error of 10\%. 
\end{tablenotes}
\end{threeparttable}
}

\clearpage
\begin{threeparttable}
\centering
\caption{Parameters of the \HCO\ and \DCO\ Spectra in the Southern Ridge} 
\label{tb:hco-dco-params}
\begin{tabular}{ccccc}
\hline 
Position 
& Line & $T_{\rm peak}$/K $^{\rm a}$& $V_{\rm LSR}$/\kms \ $^{\rm b}$ & $\Delta v$/\kms \ $^{\rm c}$ \\ \hline
\multirow{2}{*}{A}
& \HCO\ (\hcoa)	& 1.15(14) & 5.14(8)  & 2.46(20) \\  
& \DCO\ (\dcoa)	& 0.29(7) & 5.3(2) & 1.9(5) \\  \hline
\multirow{2}{*}{B}
& \HCO\ (\hcoa)	& 1.56(22) & 4.41(4)  & 0.84(10)  \\  
& \DCO\ (\dcoa)	& 0.56(17) & 4.60(10)  & 0.77(24) \\  \hline
\multirow{2}{*}{C}
& \HCO\ (\hcoa)	& 2.7(5)& 4.03(2)  & 0.31(6)  \\  
& \DCO\ (\dcoa)	& 1.2(4)& 4.31(8)  & 0.37(20) \\  \hline
\multirow{2}{*}{D}
& \HCO\ (\hcoa)	& 0.51(19) & 4.7(2) & 1.6(6) \\  
& \DCO\ (\dcoa)	& $<0.20$$^{\rm d}$ & - & - \\  \hline
\end{tabular}
\begin{tablenotes}
\footnotesize
\item 
The numbers in parentheses represent the error in units of the last significant digits, 
which is derived from the 2$\sigma$ noise error and the intensity calibration error of 10\%. 
\item[$^{\rm a}$] Peak intensity {\bff derived by the Gaussian fitting for each spectrum.}  
\item[$^{\rm b}$] LSR velocity at the intensity peak {\bff derived by the Gaussian fitting for each spectrum.} 
\item[$^{\rm c}$] FWHM of the line width {\bff derived by the Gaussian fitting for each spectrum.}   
\item[$^{\rm d}$] 
{\bff Since the \DCO\ emission is too weak to perform the Gaussian fitting at Position~D, 
we employ the 2$\sigma$ value as the upper limit for $T_{\rm peak}$.} 
\end{tablenotes}
\end{threeparttable}

\clearpage
\begin{threeparttable}
\centering
\caption{Deuterium Fractionation Ratio in the Southern Ridge} 
\label{tb:DFR}
\begin{tabular}{cccc}
\hline 
Position 
& $N$(H$^{13}$CO$^+$)/$10^{11}$ cm$^{-2}$ &  $N$(DCO$^+$)/$N$(H$^{13}$CO$^+$) &  $N$(DCO$^+$)/$N$(HCO$^+$)  \\ \hline  
A & $12^{+11}_{-3}$	& $0.3^{+0.2}_{-0.1}$	& $0.005^{+0.003}_{-0.002}$ \\  
B & $7^{+25}_{-2}$	& $0.4^{+0.2}_{-0.3}$	& $0.007^{+0.003}_{-0.005}$ \\  
C & $5^{+10}_{-1} $	& $0.5^{+0.3}_{-0.4}$	& $0.009^{+0.005}_{-0.006}$ \\  
D & $6^{+57}_{-3} $	& $<0.94$	& $<0.015$ \\\hline
\end{tabular}
\begin{tablenotes}
\footnotesize
\item 
The temperature and density ranges are assumed to be equal to that derived from the non-LTE analysis of the SO lines (Table~\ref{tb:temp-abundance_SO}). 
\end{tablenotes}
\end{threeparttable}

\clearpage
\centering 
\begin{figure}
\iffigure
\epsscale{1.0}
\includegraphics[bb = 0 0 100 450, scale = 1.0]{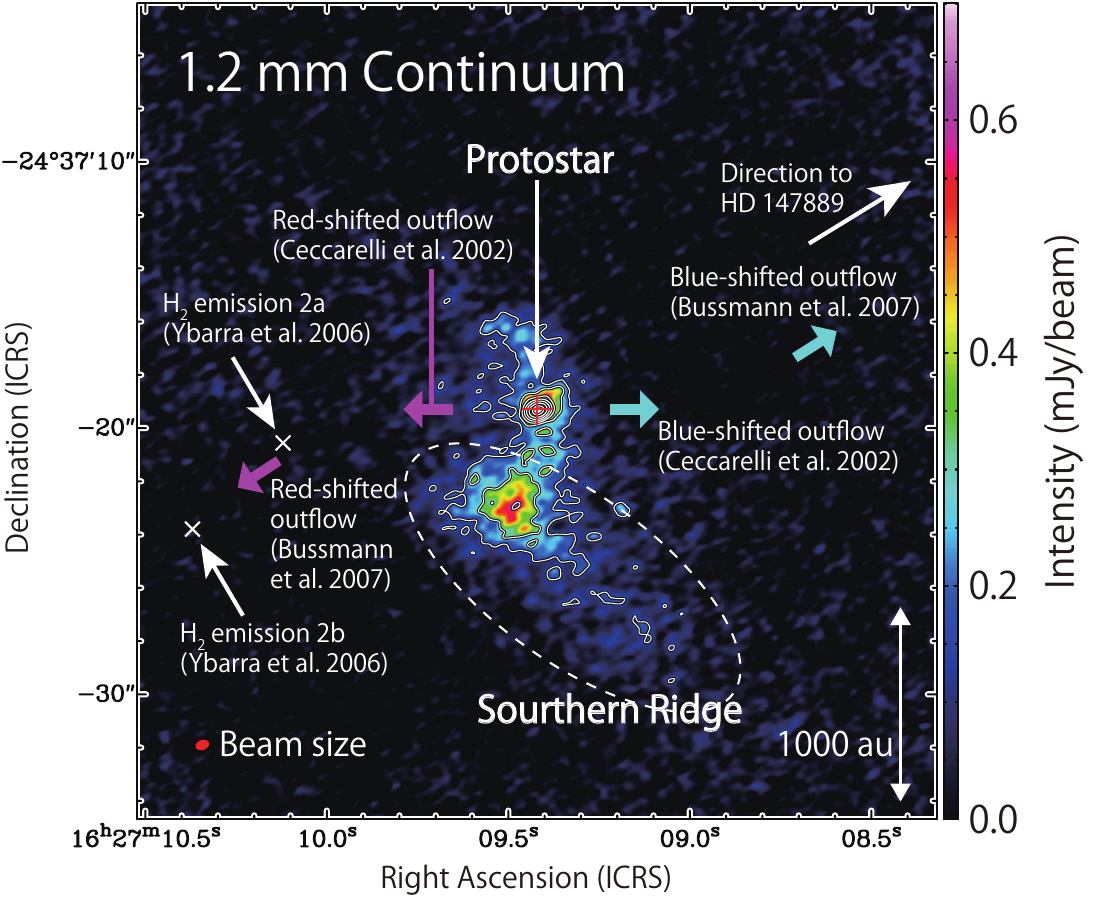} 
\fi 
\caption{%
1.2 mm continuum map. 
The red cross represents the continuum peak position at the location of the Elias 29 protostar, 
where the peak intensity is 15 \mjypb. 
The southern ridge component is indicated by the white dashed ellipse. 
Contour levels are every two times starting from 5$\sigma$ 
(i.e. 5$\sigma$, 10$\sigma$, 20$\sigma$, ...), 
where \noteRMS{30~\ujypb}. 
The beam size is shown in the red ellipse at the bottom left corner of the map.
Magenta and cyan arrows represent 
the directions of the red-shifted and blue-shifted outflow lobes, respectively, 
previously reported by \citet{Ceccarelli et al.(2002)} and \citet{Bussmann et al.(2007)}. 
{\bff White crosses represent the positions 
where the \hydro\ emission was reported by \citet{Ybarra et al.(2006)}. 
White arrow on the northwestern side of the panel represents 
the direction to the B-type star \Bstar\ taken from \citet{Rocha & Pilling(2018)}.} 
} 
\label{fig:continuum}
\end{figure}

\clearpage
\centering
\begin{sidewaysfigure}
\iffigure
\epsscale{1.0}
\includegraphics[bb = -50 0 100 1200, scale = 0.35]{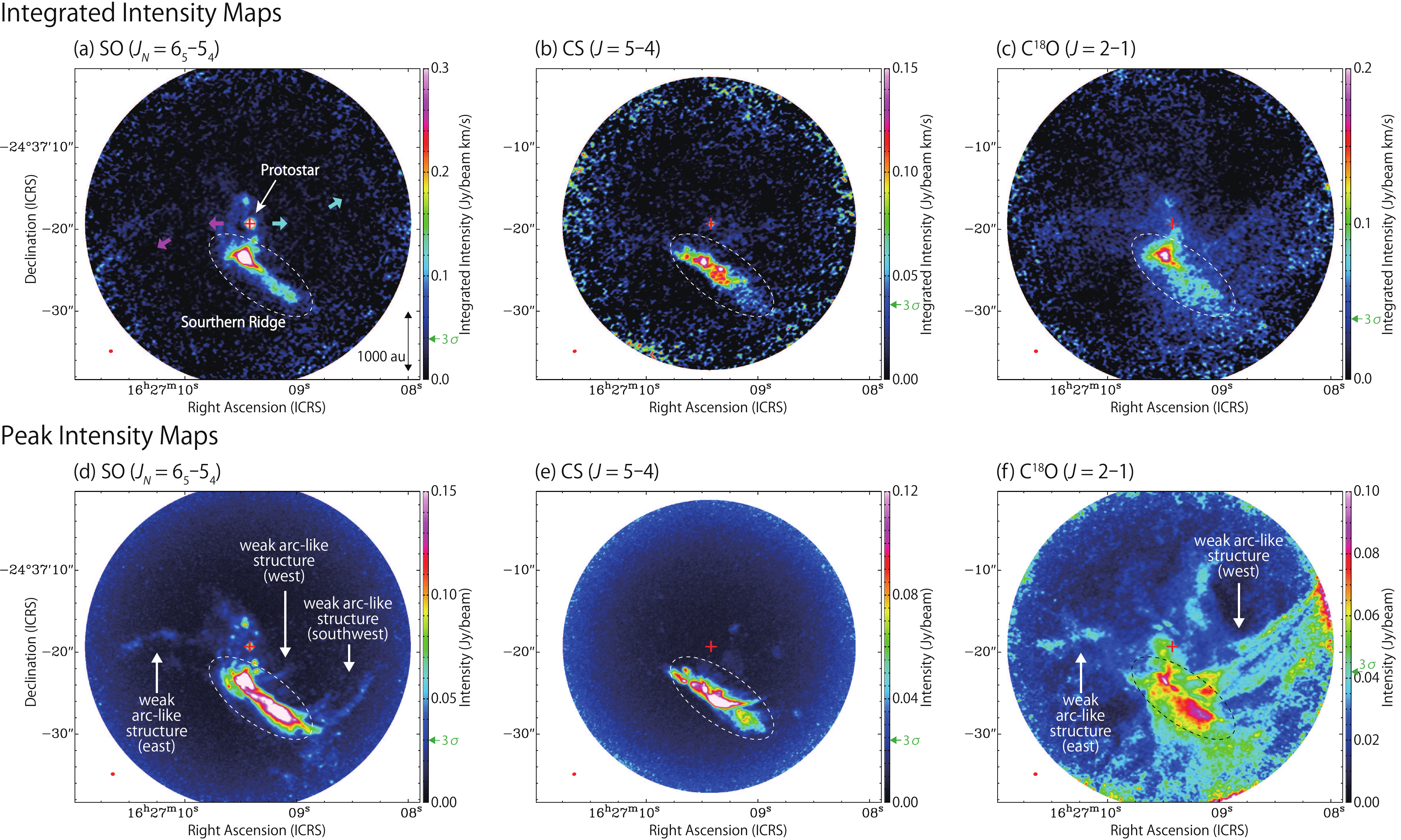} 
\fi
\caption{%
(a,b,c) Integrated intensity (moment 0) maps of SO (\soa), CS ($J=5-4$), and \co\ ($J=2-1$), respectively. 
The velocity range for the integration is from $-20$ to 30 \kms. 
(d, e, f) Peak intensity (moment 8) maps of the three molecular lines. 
The velocity range taken into account is from $-20$ to 30 \kms. 
The red cross in each panel represents the continuum peak position: \contpeakposition. 
The red ellipse at the bottom left corner in each panel represents the beam size for each molecular line. 
The magenta and cyan arrows in the top left panel are 
the same as those in Figure~\ref{fig:continuum}. 
{\bff White and black dotted ellipses indicate 
the southern ridge component. 
Green arrow indicates the 3$\sigma$ noise level in the color bar.} 
}
\label{fig:mom0-8}
\end{sidewaysfigure}

\clearpage
\centering
\begin{sidewaysfigure}
\iffigure 
\epsscale{1.2}
\includegraphics[bb = 0 0 100 2300, scale = 0.17]{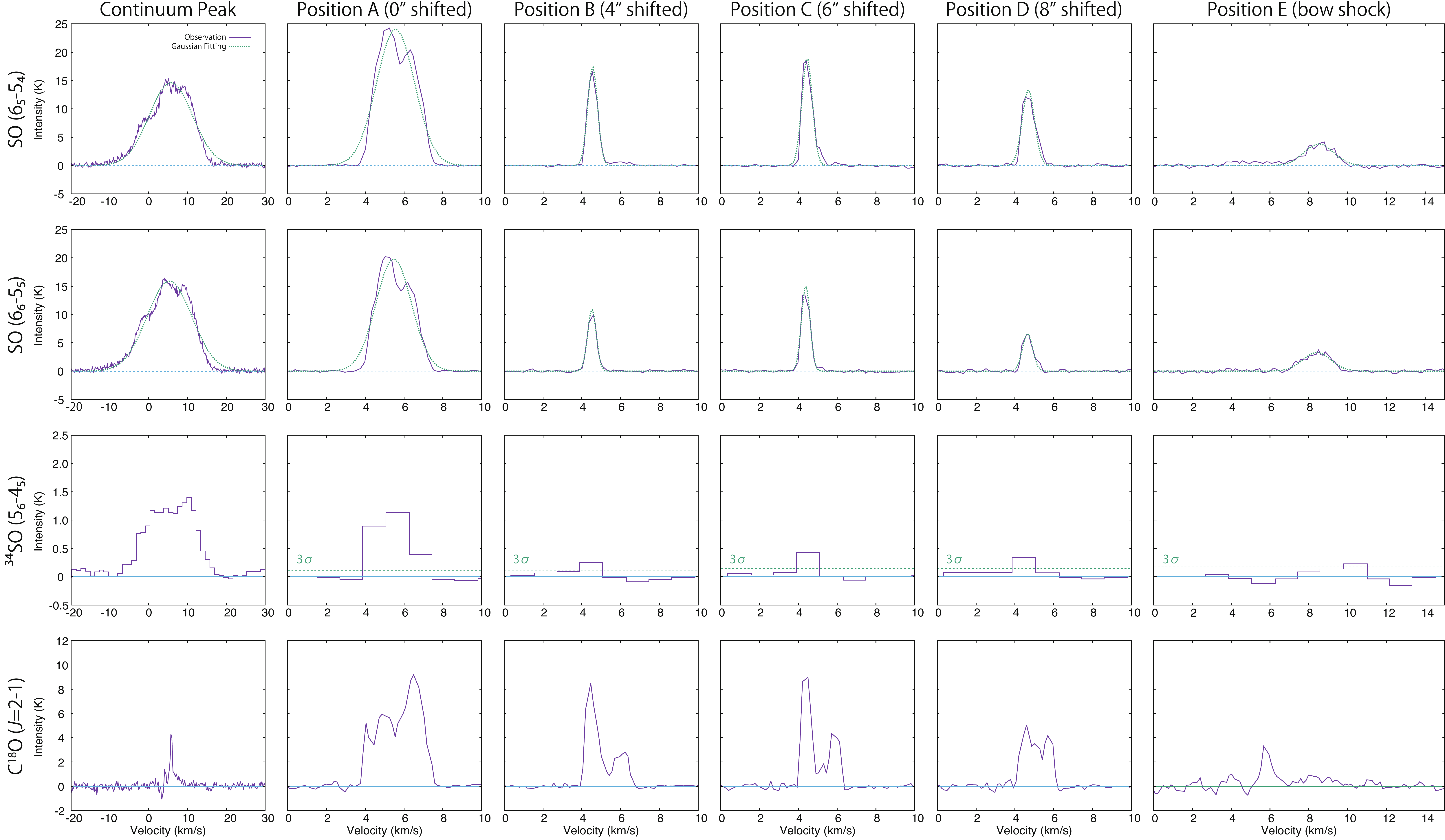} 
\fi
\caption{%
Spectra of the SO (\soa, \sob) and \SO\ (\thsoa) lines at \posAE\ (Figure~\ref{fig:ratio-Trot_SO}). 
\noteCoordsAEwithoutFig. 
Each spectrum is obtained as the average over the circular region with a diameter of 1\arcsec\ centered at each of \posAD, 
while the spectrum at one pixel is employed for Position~E. 
{\bff A frequency resolution of \SO\ (\thsoa) data is 1.129 MHz, 
while that of the other molecular line data is 141 kHz.} 
}
\label{fig:spectra-SO}
\end{sidewaysfigure}

\clearpage
\centering
\begin{sidewaysfigure}
\iffigure
\epsscale{1.2}
\includegraphics[bb = 0 0 100 500, scale = 0.56]{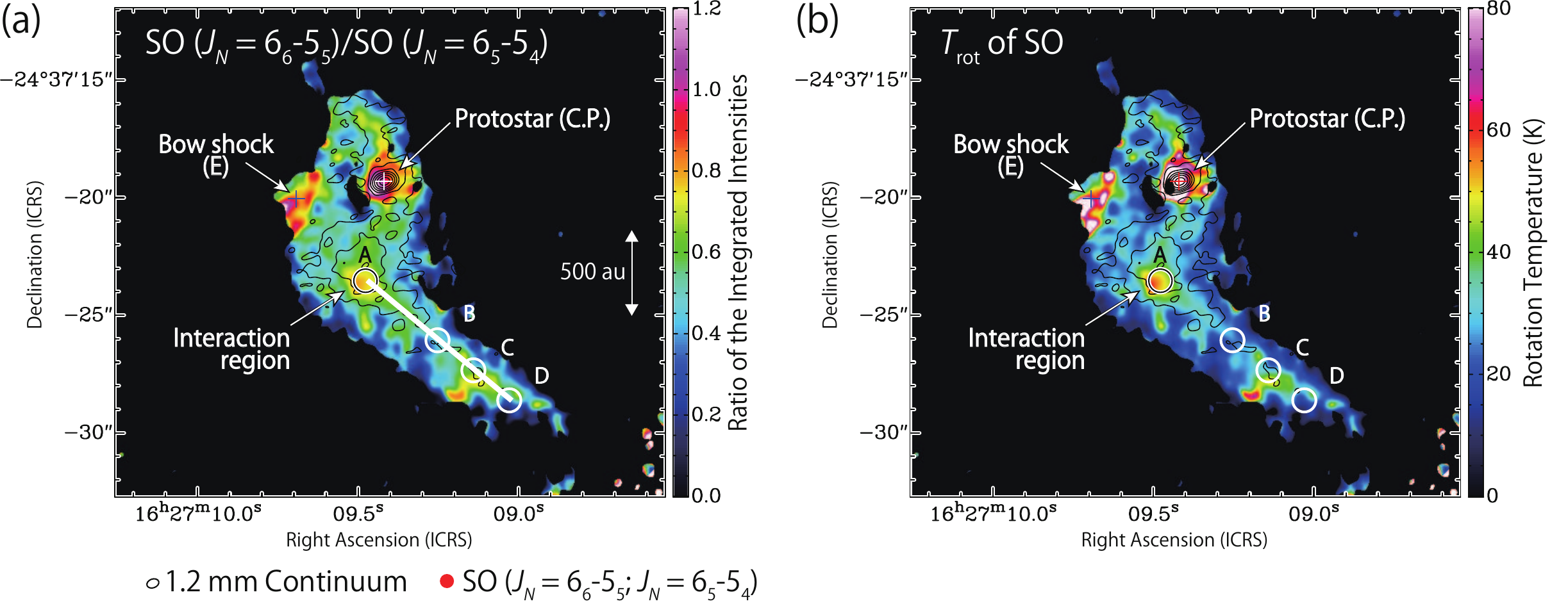} 
\fi
\caption{%
{\bff Maps of 
(a) the ratio of the integrated intensities of the two SO lines 
(b) and the rotational temperature of SO derived from the intensity ratio.
The images of the two SO lines are smoothed so that their beam sizes are to be $0\farcs6 \times 0\farcs6$, 
as shown by the red circle outside the panels. 
{\bff The velocity range for the integration is from $-15$ to $+23$~\kmps.} 
Black contours in each panel represent the 1.2 mm continuum emission, 
where the contour levels are the same as those in Figure~\ref{fig:continuum}. 
The beam size of the continuum image is shown by a black ellipse outside the panels.} 
(a) 
{\bff Ratio} of the integrated intensity of the SO (\sob) line to the SO (\soa) line, 
{\bff which corresponds to $W_2 / W_1$ in Equation~(\ref{eq:lte}).} 
The pixels where the integrated intensity of the SO (\soa) line is less than 2$\sigma$ are shown in black, 
where \noteRMS{18~\mjypb}. 
(b) 
{\bff Rotational} temperature of SO derived from the ratio of the integrated intensity of the two line emissions in the panel (a). 
The pixels with negative values are shown in black, 
where the ratio of the integrated intensity is less than 0 or higher than the high temperature limit ($1.15$ for $T_{\rm rot} = \infty$). 
%
{\bff See the footnotes for Table~\ref{tb:coords} for the explanation about Positions~A-E.} 
}
\label{fig:ratio-Trot_SO}
\end{sidewaysfigure}

\clearpage
\centering
\begin{sidewaysfigure}
\iffigure
\epsscale{1.2}
\includegraphics[bb = 0 0 100 800, scale = 0.43]{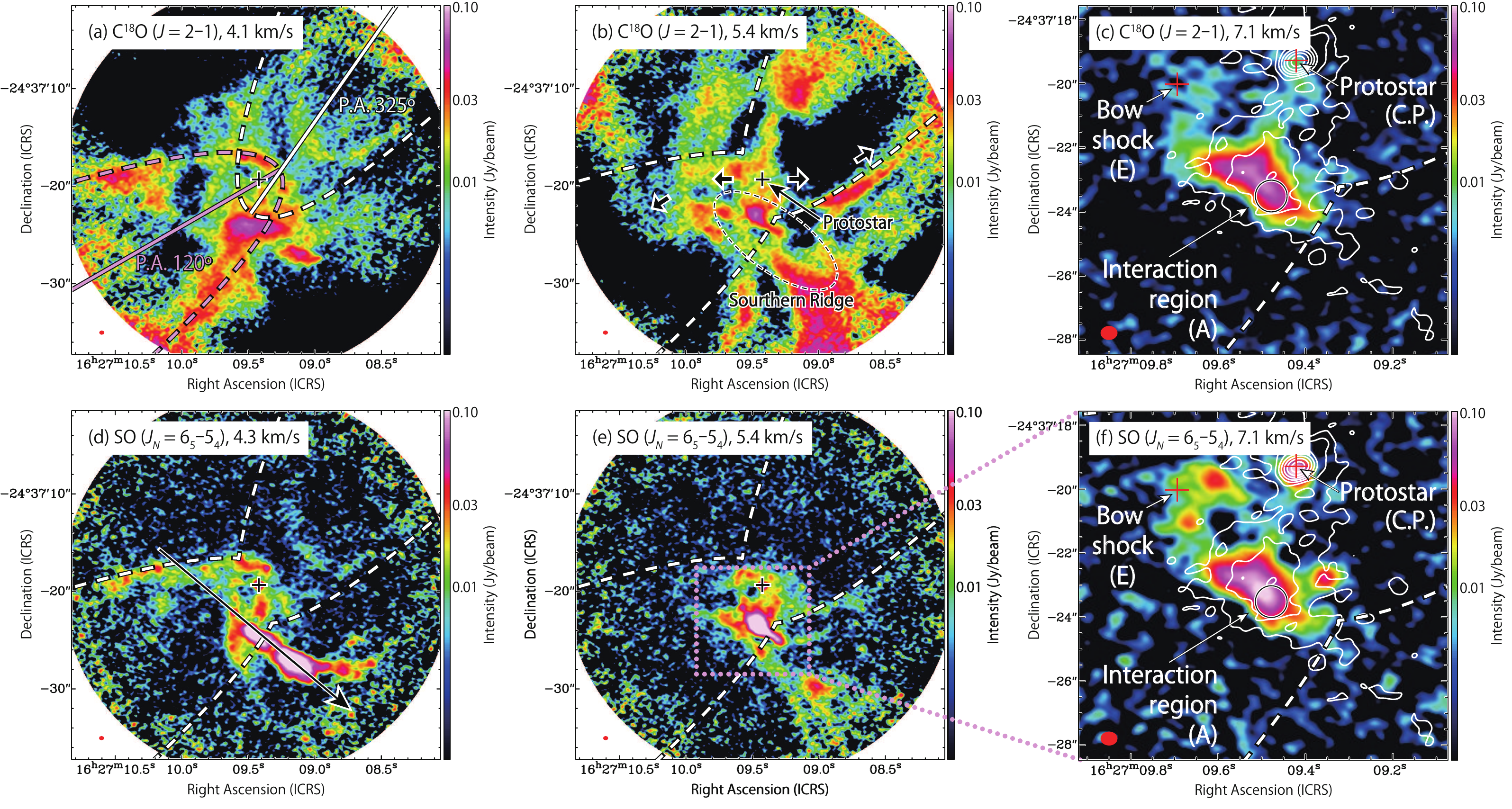} 
\fi
\caption{%
The velocity channel maps of the \CO\ (\coa; a$-$c) and SO (\soa; d$-$f) lines. 
Panels~(a,~d) show the velocity channel maps of the two molecular lines near the systemic velocity of the protostar ($\sim$$+4$~\kmps). 
Panels~(b,~e) show their velocity channel maps with the red-shifted velocity 
{\bff (5.4~\kmps).} 
{\bff Panels~(c,~f) show the maps with a more red-shifted velocity (7.1~\kmps), 
focusing on the region around Position~A. 
The pink dotted rectangle in panel~(e) indicates the region shown in panels~(c,~f). 
White contours in panels~(c,~f) represent the continuum emission, 
where the contour levels are the same as those in Figure~\ref{fig:continuum}.} 
The beam size is depicted by a red ellipse at the bottom left corner in each panel. 
{\bff Black or red cross in each panel represents the protostellar position (C.P.).} 
{\bff White and pink dashed lines indicate the outflow morphology 
obtained based on these panels.} 
{\bff The central axes of the parabolic shapes 
are represented by white and pink solid lines in panel~(a), 
which have the position angles of 120\degr\ and 325\degr, respectively.} 
The directions of the outflow lobes previously reported 
are indicated by the black arrows in panel~(c), 
which are the same as the magenta and cyan arrows in Figure~\ref{fig:continuum}. 
Black arrow in panel~(d) indicates 
the position axis along which Figure~\ref{fig:SO-southernridge-PV} is obtained. 
}
\label{fig:mom0-outflow}
\end{sidewaysfigure}

\clearpage
\centering
\begin{figure}
\iffigure
\epsscale{1.0}
\includegraphics[bb = -50 0 100 500, scale = 1.0]{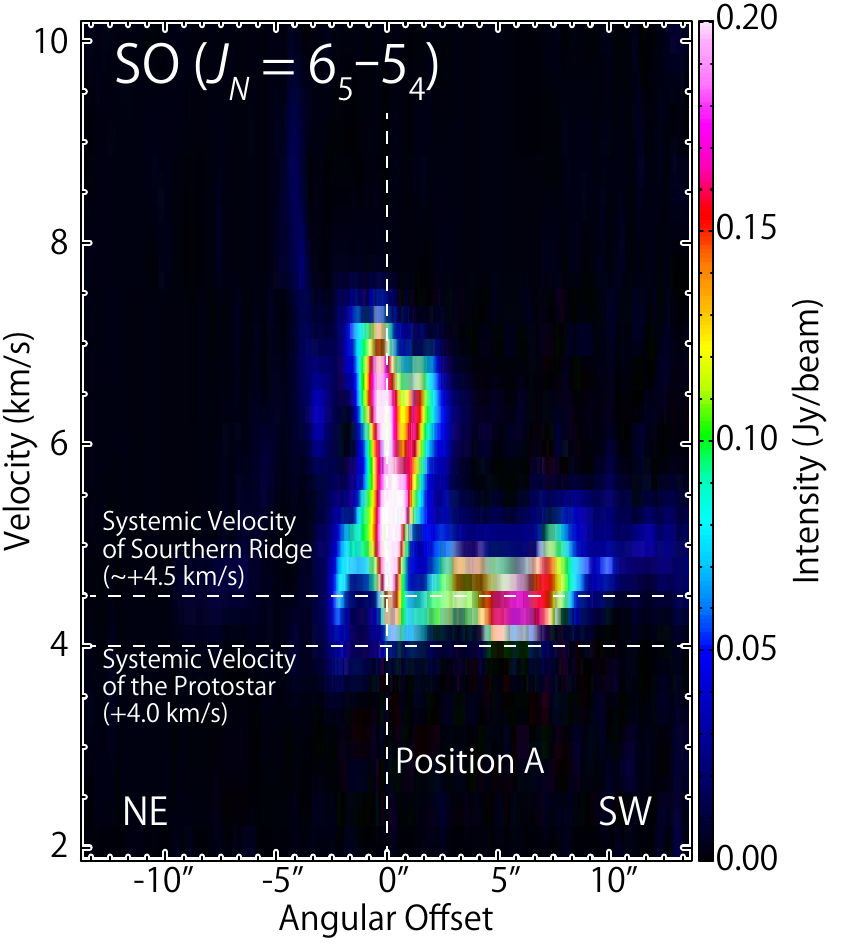} 
\fi
\caption{%
Position-Velocity diagram of the SO (\soa) line. 
The position axis is taken along the southern ridge component 
as shown by the black arrow in Figure~\ref{fig:mom0-outflow}(d). 
Angular offset of 0\arcsec\ shown by the dashed vertical line is taken at Position~A. 
The dashed horizontal lines represent the systemic velocity of the southern ridge 
{\bff ($+4.5$~\kmps)} 
and that of the protostar ($+4.0$~\kmps). 
}
\label{fig:SO-southernridge-PV}
\end{figure}

\clearpage
\centering
\begin{sidewaysfigure}
\iffigure
\epsscale{1.2}
\includegraphics[bb = 0 0 100 500, scale = 0.45]{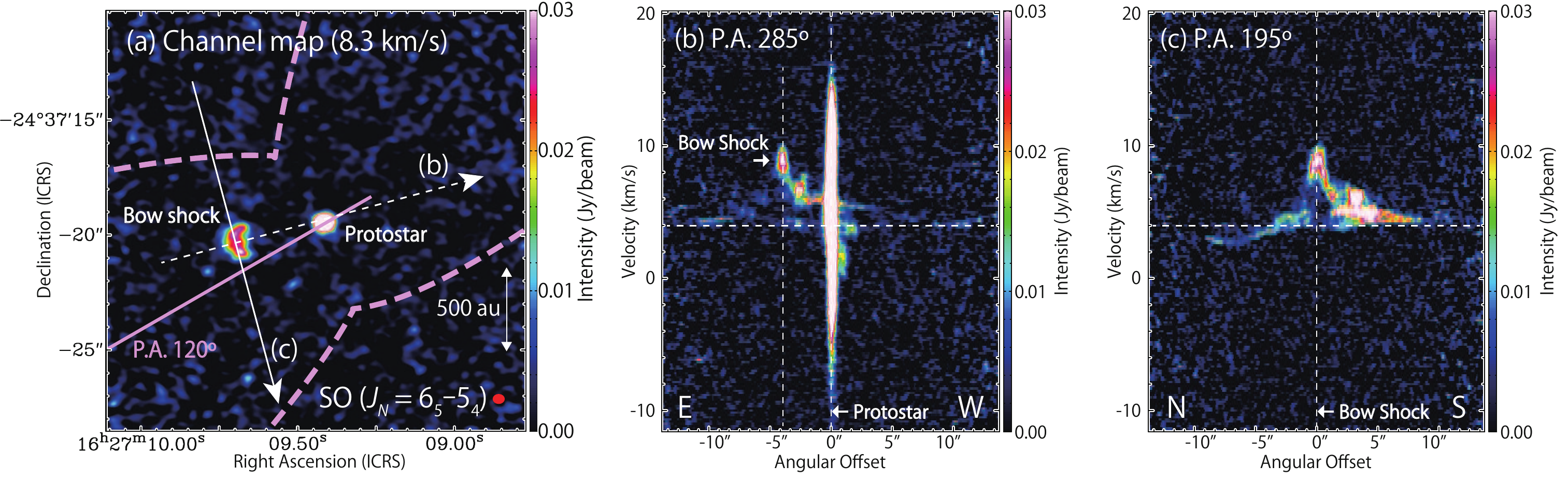} 
\fi
\caption{%
(a) Velocity channel map of the SO (\soa) line. 
Velocity of $+8.3$~\kmps\ is red-shifted from the systemic velocity of the protostar ($\sim +4$~\kmps). 
{\bff Pink dashed and solid lines are the same as those in Figure~\ref{fig:mom0-outflow}.} 
(b, c) Position-Velocity diagrams of the SO (\soa) line. 
{\bff Position axes for panels~(b,~c) are shown by the white dashed and solid arrow in panel~(a), respectively. 
White vertical lines {\bff in panels~(b,~c)} represent the protostellar position and Position~E, {\bff respectively.}} 
White horizontal lines in panels~(b,~c) represent the systemic velocity of the protostar ($+4.0$~\kmps). 
}
\label{fig:bowshock}
\end{sidewaysfigure}

\clearpage
\centering
\begin{sidewaysfigure}
\iffigure
\epsscale{1.2}
\includegraphics[bb = 0 0 100 1000, scale = 0.23]{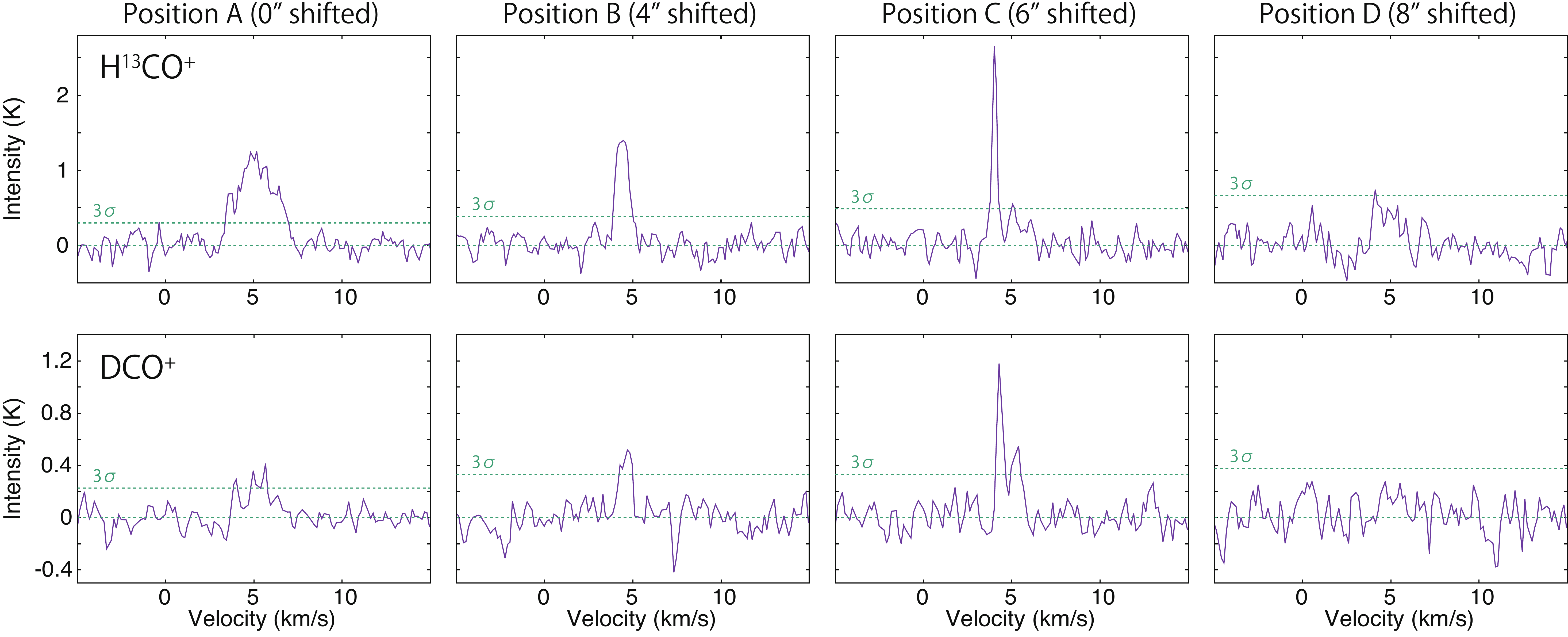} 
\fi
\caption{%
Spectra of the \HCO\ (\hcoa) and \DCO\ (\dcoa) lines. 
The spectra are obtained for \posAD\ (Figure~\ref{fig:ratio-Trot_SO}). 
\noteCoordsADwithoutFig. 
Each spectrum represents the average over the circular region with a diameter of 1\arcsec\ centered at each of \posAD. 
}
\label{fig:H13COp-DCOp-spectra}
\end{sidewaysfigure}

\end{document}